\documentclass[a4paper,11pt]{article}
\usepackage[dvips]{graphicx}
\usepackage{hyperref}
\usepackage{latexsym}
\usepackage{amssymb}
\usepackage{amsmath}
\usepackage{fancyhdr}
\usepackage{multirow}

\newcommand{\oo}{\mathcal{O}}
\newcommand{\lat}{\textrm{lat}}
\newcommand{\free}{\textrm{free}}
\newcommand{\cont}{\textrm{cont}}
\newcommand{\MSb}{\overline{\textrm{MS}}}
\newcommand{\MSt}{\widetilde{\textrm{MS}}}

\newcommand{\Oag}{\mathcal{O}(a^2g^2)}
\newcommand{\Lqcd}{\Lambda_{\rm QCD}}

\begin{document}

\begin{flushright}
 DESY 12-118\\
 SFB/CPP-12-44
\end{flushright}

\begin{center}
\Large
Non-perturbative renormalization in coordinate space for $N_f=2$ maximally
twisted mass fermions with tree-level Symanzik improved gauge action

\normalsize

\vspace{0.6cm}

Krzysztof Cichy$^{1,2}$, Karl Jansen$^{1}$, Piotr Korcyl$^{1,3}$\\
\vspace{0.3cm}
$^{1}$\emph{NIC, DESY, Platanenallee 6, D-15738 Zeuthen, Germany}\\
$^2$ \emph{Adam Mickiewicz University, Faculty of Physics,\\
Umultowska 85, 61-614 Poznan, Poland}\\
$^3$ \emph{M. Smoluchowski Institute of Physics, Jagiellonian University,\\
Reymonta 4, 30-059 Krakow, Poland}\\

%\vspace{0.3cm}

\begin{center}
\includegraphics
[width=0.2\textwidth,angle=0]
{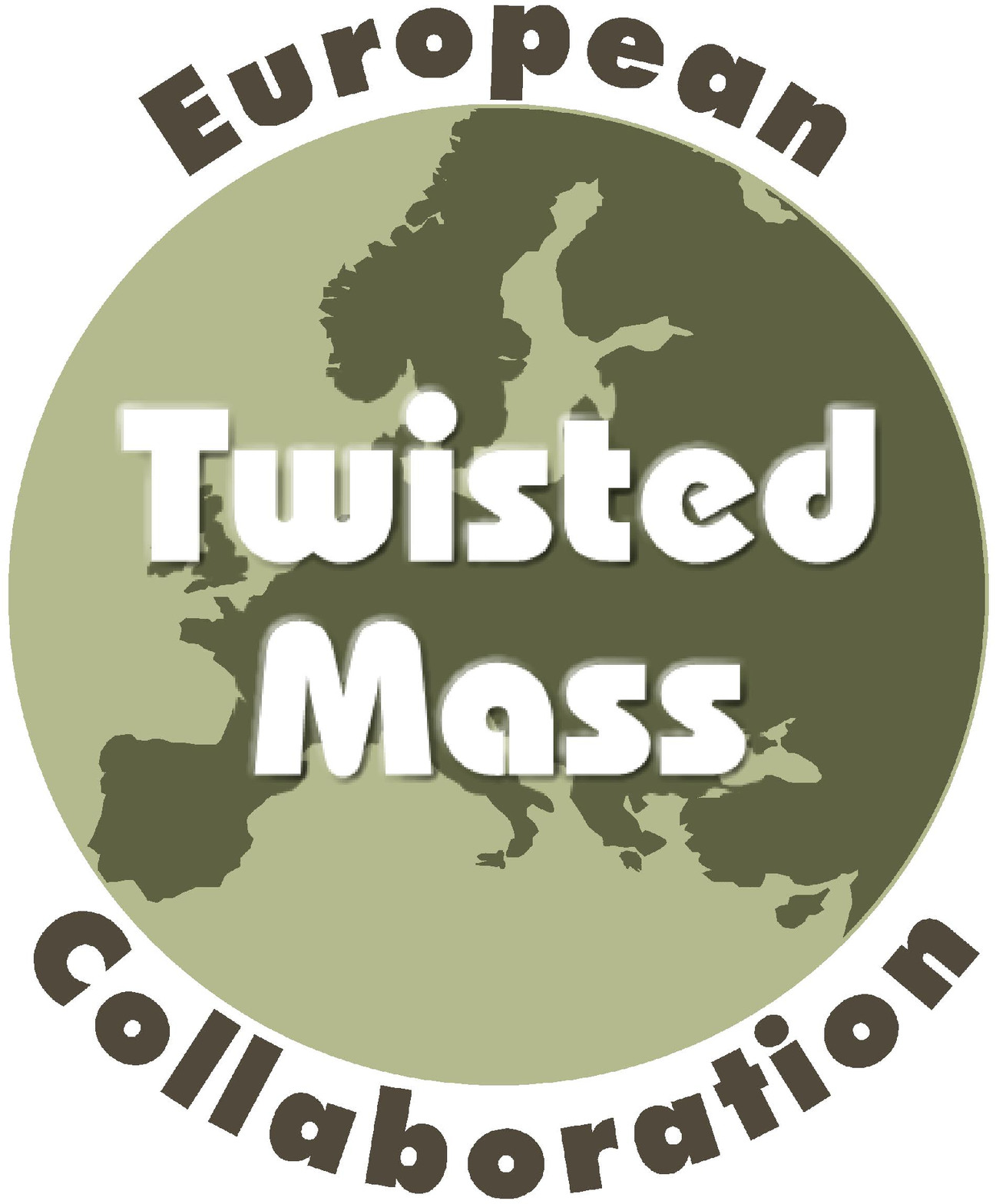}
\end{center}

\begin{abstract}
\noindent We present results of a lattice QCD application of a coordinate space renormalization scheme
for the extraction of renormalization constants for flavour non-singlet bilinear quark operators. The
method consists in the analysis of the small-distance behaviour of correlation functions in
Euclidean space and has several theoretical and practical advantages, in particular:
it is gauge invariant, easy to implement and has relatively low computational cost. The
values of renormalization constants in the X-space scheme can be converted to the $\MSb$
scheme via 4-loop continuum perturbative formulae. Our results for $N_f=2$ maximally twisted mass
fermions with tree-level Symanzik improved gauge action are compared to the ones from the RI-MOM
scheme and show full agreement with this method.
\end{abstract}


\end{center}

\section{Introduction}
\label{sec. intro}

For many physical quantities to be computed in Lattice QCD, 
renormalization is an essential ingredient.
Therefore, it is extremely important to have, ideally, 
several different methods available 
for the determination of renormalization constants (RCs)
that are moreover 
non-perturbative.
In this paper, our main aim is to show that the extraction of RCs from the behaviour of
correlation functions in coordinate space (termed \emph{X-space method} below) is 
feasible and, besides providing the RCs in a gauge invariant manner, 
has certain advantages. 
Therefore, the X-space method has the potential 
as an useful alternative to other widely used
methods, although, as we will show in this paper, further 
improvements are still necessary to allow for a precise 
computation of RCs  
with fully controlled systematic uncertainties.

One way to extract RCs for Lattice QCD is Lattice Perturbation
Theory (LPT), see Ref.~\cite{Capitani:2002mp} for 
a comprehensive review. While with LPT  
it is difficult to go beyond 1-loop order\footnote{An example of a 2-loop
computation of RCs for clover fermions is given in
Refs.~\cite{Skouroupathis:2007jd}, \cite{Skouroupathis:2008mf}. 2-loop LPT computation of quark
mass renormalization is presented in Ref.~\cite{Mason:2005de}. 2-loop relation
between the bare lattice coupling and the MS coupling in pure SU(N) gauge theories can be found
in Refs.~\cite{Luscher:1995nr}, \cite{Luscher:1995np}.}, 
it is still very useful, if not necessary
%Therefore, nowadays it is usually
to complement 
non-perturbative computations of RCs.
One of developments in LPT is  
a tadpole
resummation method proposed in Refs.~\cite{Parisi:1980pe}, \cite{Lepage:1992xa}, which has been 
further explored in Refs.~\cite{Panagopoulos:1998vc}, \cite{Constantinou:2006hz}
to encompass effects of various gluonic and fermionic discretizations. 
Tadpole resummation amounts to a redefinition
of the coupling constant. 
A method to go beyond the 1-loop (or sometimes 2-loop) approximation  
is the framework of Numerical Stochastic Perturbation Theory
(NSPT) \cite{DiRenzo:1994av}, \cite{DiRenzo:1994sy}, \cite{Di Renzo:2004ge}. This method,
originates from the Stochastic Quantization theory of Parisi and Wu \cite{Parisi:1980ys},
\cite{Damgaard:1987rr}, which can be used to formulate Stochastic Perturbation Theory, 
which is then applied  
to stochastically evaluate perturbatively accessible quantities, such 
as RCs or the running coupling constant by numerical 
methods. 
This makes it possible to go to high loop orders in LPT. 
The method has been successfully applied in the context of renormalization, see e.g. Refs.~\cite{Di
Renzo:2006wd}, \cite{Brambilla:2010qb}, \cite{Horsley:2012ra}.

Of course, eventually a non-perturbative approach to determine 
RCs is desirable. 
The most natural approach is the use of Ward identities. 
Such an approach allows to obtain a very good precision for  
a selected number of RCs, see 
e.g. Refs.~\cite{Bochicchio:1985xa}, \cite{Jansen:2005kk}, \cite{Becirevic:2006ii},
\cite{Constantinou:2010gr}, 
for an introduction see Ref.~\cite{Vladikas:2011bp}.
However, the application of Ward identities is rather limited and cannot be
used for a number of important renormalization constants. Therefore, 
in order to compute a broad range of RCs 
one has to resort to different but still non-perturbative methods.

The two most widely used non-perturbative 
methods are the Regularization-Independent Momentum Subtraction (RI-MOM)
and the Schr\"odinger Functional (SF) renormalization schemes.
For an extensive review of their practical usage, see Refs.~\cite{Sommer:2002en},
\cite{Aoki:2010yq}.
The RI-MOM method \cite{Martinelli:1994ty}, also called the
Rome-Southampton method, is similar in the definitions of renormalization conditions to continuum
perturbation theory. It consists in the lattice evaluation of quark propagators of the relevant
operators between external quark and gluon states and requires 
hence to fix the gauge. 
Since the original method has been proposed, it was subject to many improvements of its reliability,
including the perturbative computation of the relevant $\Oag$ discretization effects
\cite{Becirevic:2004ny}, \cite{Constantinou:2009tr} and Goldstone pole subtraction
\cite{Cudell:1998ic}, \cite{Giusti:2000jr}.
The RI-MOM approach has been applied in several lattice setups. Since 
in the end we want to compare our results for the RCs with the ones 
computed in the RI-MOM scheme using configurations generated
by the European Twisted Mass Collaboration (ETMC) \cite{Boucaud:2007uk}, \cite{Boucaud:2008xu},  
we mention here the work related to this setup, i.e. Refs.~\cite{Dimopoulos:2007fn},
\cite{Constantinou:2010gr}, \cite{Alexandrou:2012mt}.
Other lattice setups included e.g. $N_f=2+1+1$ MTM
fermions on Iwasaki gauge action \cite{ETM:2011aa} and $N_f=2$ clover fermions on Wilson plaquette
action \cite{Gockeler:2010yr}. 
For completeness, we note in passing the approach of the 
spectral projectors method, see Ref.~\cite{Giusti:2008vb}, 
which can be used to compute the 
scale independent ratio $Z_P/Z_S$ and which has already been applied to 
twisted mass fermions. 

The other widely used non-perturbative method, the Schr\"odinger functional \cite{Luscher:1992an},
\cite{Sint:1993un} is defined in a four-dimensional Euclidean space with Dirichlet boundary
conditions in time direction.
Combined with finite-size techniques \cite{Luscher:1991wu}, it can be used to formulate a
SF renormalization scheme \cite{Jansen:1995ck}, with the spatial lattice size used to set the
(inverse of) renormalization scale. 
In the SF scheme, the scale dependence and the continuum limit
are separated by determining the continuum limit of the lattice step scaling function first, 
before the scale dependence is considered. 
The SF scheme is formulated in a completely gauge invariant manner, 
however, carrying through the corresponding renormalization programme 
bears some complexity. 
The method has been successfully applied e.g. in Refs.~\cite{Luscher:1996jn},
\cite{Capitani:1998mq}, for an introduction see e.g. Refs.~\cite{Sommer:1997xw},
\cite{Sommer:2006sj}. 
Concerning maximally twisted mass fermions as we will address here, 
the SF scheme formalism has been developed in Refs.~\cite{Sint:2005qz}, \cite{Sint:2010eh}, 
leading to the introduction of chirally rotated boundary conditions.  
For first applications, however 
in the quenched theory only,  
we refer to Refs.~\cite{Lopez:2008ns},
\cite{Lopez:2009yc}, \cite{Sint:2010xy}. 

Our present paper concerns yet another approach to non-perturbative renormalization. The X-space
method was suggested in Ref.~\cite{Martinelli:1997zc} and the results of its first
implementation were presented in Refs.~\cite{Becirevic:2002yv}, \cite{Gimenez:2003rt},
\cite{Gimenez:2004me}. It consists in imposing a renormalization condition directly on
gauge invariant correlation functions at small Euclidean distances, leading thus to 
the advantage of a
gauge invariant determination of RCs. A further advantage is 
the absence of contact terms. 
In principle, this makes the X-space method also suitable for renormalization of four-fermion operators,
relevant e.g. for the computations of weak matrix elements, although in this paper we limit
ourselves to RCs of bilinear quark operators. 
For the X-space method, the natural
limitation is the necessity of existence of a renormalization window, namely the regime of
Euclidean distances large enough that uncontrollable discretization effects can be avoided, but
\emph{at the same time} small enough to make contact with (continuum) perturbation theory (i.e.
much smaller than the inverse of $\Lqcd$). 
The purpose of this paper is to explore the X-space method in practice
and in particular to see, whether in current simulations, taking $N_f=2$ 
LQCD setup as an example, the above mentioned window exists and whether 
the method is feasible. By comparing to results for RCs using 
the RI-MOM scheme for the $N_f=2$ maximally twisted mass setup, 
we can also check the validity of the approach. We will also make an attempt
to determine the systematic errors appearing in the computation 
of RCs with the X-space method, which will allow us in the end
to address its competitiveness with the methods described above.

The paper is organized as follows. In Sec.~\ref{sec. theory}, we present the theoretical
foundations of the method. Details of its practical implementation  are given in
Sec.~\ref{sec. analysis}, whereas our final results are presented in Sec.~\ref{sec. analysis2}.
Sec.~\ref{sec. conclusions} concludes and discusses the prospects for
broader application of the method.

\section{Theoretical foundations}
\label{sec. theory}

In this section, we briefly describe theoretical aspects of the X-space method and our lattice
setup. We start by summarizing twisted mass Lattice QCD,
then we introduce the correlation functions whose numerical analysis will be presented in
Secs.~\ref{sec. analysis}, \ref{sec. analysis2}. Next, we provide the definition of the X-space
renormalization scheme and
discuss some of its properties. Finally, we quote some results from continuum perturbation theory
which are needed for the conversion of our results to the $\MSb$ scheme and to the evolution of the
renormalization constant to the scale of 2 GeV.

\subsection{Twisted mass Lattice QCD}
\label{sec. tmlqcd}

We consider correlation functions calculated from gauge link configurations generated with $N_f=2$
dynamical quarks by ETMC \cite{Boucaud:2007uk}, \cite{Boucaud:2008xu}. The action $S[\psi,
\bar{\psi}, U]$,
$S[\psi, \bar{\psi}, U] = S_G[U] + S_F[\psi, \bar{\psi},U]$ used to generate these configurations is
given by the tree-level Symanzik improved gauge action \cite{Weisz:1982zw}
\begin{equation}
 S_G[U] = \frac{\beta}{6}\Big( b_0 \sum_{x,\mu \ne \nu} \textrm{Tr} \big( 1 - P^{1\times 1}(x;\mu,\nu) \big) 
+ b_1 \sum_{x,\mu \ne \nu} \textrm{Tr}\big( 1 - P^{1 \times 2}(x; \mu, \nu) \big) \Big),
\end{equation}
with $\beta=6/g^2$, $g$ -- bare coupling, $b_0 = \frac{5}{3}$ and $b_1 = -\frac{1}{12}$, $P^{1\times
1}$, $P^{1\times 2}$ denote the plaquette and the rectangular Wilson loop, respectively; the
Wilson twisted mass fermion action is given in the twisted basis by
\cite{Frezzotti:2000nk}, \cite{Frezzotti:2003ni}, \cite{Frezzotti:2004wz}, \cite{Shindler:2007vp}
\begin{equation}
 S_F[\psi, \bar{\psi}, U] = a^4 \sum_x \bar{\psi}(x) \big( D_W + i \mu_q \gamma_5 \tau_3 \big) \psi(x),
 \label{eq. twisted fermion action}
\end{equation}
with
\begin{equation}
 D_W = \frac{1}{2} \big( \gamma_{\mu} (\nabla_{\mu} + \nabla^*_{\mu}) - a \nabla^*_{\mu} \nabla_{\mu} \big) + m_0.
\end{equation}
As usual, in the above formulae $a$ denotes the lattice spacing, whereas $\nabla_{\mu}$ and $\nabla^*_{\mu}$ are the discretized 
gauge covariant forward and backward derivatives. The bare untwisted and twisted fermion masses are
denoted by $m_0$ and $\mu_q$, respectively. $\psi(x)$ is a two-component vector in flavour space
build of two spinors $u(x)$ and $d(x)$. The matrix $\tau^3$ in Eq.~\eqref{eq. twisted fermion
action} acts in this flavour space. Such formulation allows for an automatic $\mathcal{O}(a)$
improvement of physical observables, provided the hopping parameter $\kappa = (8+2 a m_0)^{-1}$ is
tuned to maximal twist by setting it to its critical value, at which the PCAC quark mass vanishes
\cite{Frezzotti:2000nk}, \cite{Frezzotti:2005gi}, \cite{Jansen:2005kk}.

\subsection{Correlation functions}
\label{sec. functions}

Correlation functions considered in this work are constructed from flavour non-singlet bilinear
quark operators and are of the form
\begin{equation}
 C_{\Gamma \Gamma}(X) = \langle \oo_{\Gamma}(X) \oo_{\Gamma}(0)\rangle,
 \label{eq. correlation functions}
\end{equation}
where
\begin{equation}
 \oo_{\Gamma}(X) = \bar{\psi}(X) \Gamma \psi(X)
\end{equation}
with $\oo_{\Gamma} = \{S, P, V_{\mu}, A_{\mu} \}$ for $\Gamma = \{ 1, \gamma_5, \gamma_{\mu},
\gamma_{\mu} \gamma_5 \}$. $X$ denotes the Euclidean vector: $X \equiv (x_1, x_2, x_3, x_4)
\equiv (x,y,z,t)$.

In order to avoid confusion, it is important to state explicitly our
conventions of denoting the correlation functions and their corresponding RCs.
We label RCs according to the un-twisted Wilson
case notation. Therefore, by $Z_V$ and $Z_A$ we denote the RCs of the local
vector and axial-vector currents of the standard Wilson action, although in our case they
renormalize in the twisted basis the local axial-vector and polar-vector currents,
respectively
(see Tab.~\ref{tab. convention}).

\begin{table}[t!]
 \begin{center}
  \begin{tabular}{cc}
   \hline
   \hline
	Un-twisted case & Twisted case \\
   \hline
	$(A_R)_{\mu} = Z_A A_{\mu} $ & $(A_R)_{\mu} = Z_V A_{\mu} $ \\
	$(V_R)_{\mu} = Z_V V_{\mu} $ & $(V_R)_{\mu} = Z_A V_{\mu} $ \\
	$(P_R) = Z_P P  $ & $(P_R) = Z_P P $ \\
	$(S_R) = Z_S S  $ & $(S_R) = Z_S S $\\
   \hline
   \hline
  \end{tabular}
 \end{center}
\caption{Summary of our convention of naming the correlation functions and their corresponding
renormalization constants.\label{tab. convention}}
\end{table}

\subsection{X-space renormalization scheme}
\label{sec. x-scheme}

The renormalization constants are defined non-perturbatively by the following condition
\cite{Gimenez:2004me} imposed \emph{in position space} and \emph{in the chiral limit}:
\footnote{The renormalization point will always be denoted by $X_0 = ( x_0, y_0, z_0, t_0 )$. By
$X_0^2$ we mean $X_0^2 = (x_0)^2 + (y_0)^2 + (z_0)^2 + (t_0)^2$. When the meaning of $X_0$ is
unambiguous, we will also use it to denote $X_0 = \sqrt{(x_0)^2 + (y_0)^2 + (z_0)^2 + (t_0)^2}$. }
\begin{equation}
 \lim_{a\rightarrow 0}\langle \oo^X_{\Gamma}(X) \oo^X_{\Gamma} (0) \rangle \big|_{X^2=X_0^2} =
\langle \oo_{\Gamma}(X_0) \oo_{\Gamma}(0) \rangle^{\free}_{\cont}
\label{eq. condition}
\end{equation}
where the renormalized operator is 
\begin{equation}
 \oo^X_{\Gamma}(X, X_0) = Z^X_{\Gamma}\big(X_0^2, (\hat{x}_0,\hat{y}_0,\hat{z}_0,\hat{t}_0),a\big)
\oo_{\Gamma}(X),
\end{equation}
and $X_0$ is the renormalization point. The vector $(\hat{x}_0,\hat{y}_0,\hat{z}_0,\hat{t}_0)$ is
simply given by $\frac{1}{X_0}(x_0,y_0,z_0,t_0)$ and denotes the direction of the renormalization
point\footnote{In a practical application of this method, we will use an empirical criterion (so
called ``democratic'' points) to select such direction $(\hat{x}_0,\hat{y}_0,\hat{z}_0,\hat{t}_0)$
that is least affected by lattice artefacts and average over few neighbouring directions in order to
obtain a stronger signal. Details of our procedure will be described in Sec.~\ref{sec. analysis}. A
systematic analysis of this aspect will be presented elsewhere.}.
$X_0$ must satisfy the condition $a \ll X_0 \ll
\Lambda^{-1}_{\textrm{QCD}}$ so that
the discretization effects and the connection to continuum perturbation 
theory are under control.

The condition of Eq.~\eqref{eq. condition} should be understood in the following way: For every
finite value of the lattice spacing $a$, we impose the condition 
\begin{equation}
\Big( Z^X_{\Gamma}\big(X_0^2, (\hat{x}_0,\hat{y}_0,\hat{z}_0,\hat{t}_0), a\big) \Big)^2 \langle
\oo_{\Gamma}(X_0) \oo_{\Gamma} (0) \rangle(a) = \langle \oo_{\Gamma}(X_0) \oo_{\Gamma}(0)
\rangle^{\free, \textrm{massless}}_{\cont},
\end{equation}
from which the RC at this value of $a$, 
$Z^X_{\Gamma}(X_0,(\hat{x}_0,\hat{y}_0,\hat{z}_0,\hat{t}_0), a)$,
can be calculated. The continuum limit of Eq.~\eqref{eq. condition} is then trivially satisfied.
Such procedure then guarantees that we recover the correctly normalized quantities of the continuum
theory. In order to illustrate this last statement, let us consider the isovector current
$V^a_{\mu}$. In the continuum theory it is a conserved quantity defined by the symmetry
transformations 
\begin{align}
\delta \psi &= \frac{i}{2} \alpha^a \tau^a \psi, \nonumber \\
\delta \bar{\psi} &= - \bar{\psi} \frac{i}{2} \alpha^a \tau^a. 
\label{eq. symmetry}
\end{align}
On the lattice, two discretizations are commonly used, which have the same naive continuum limit,
namely, (see e.g. Ref.~\cite{Bochicchio:1985xa})
\begin{align}
 V_{\mu}^a(X) &= \frac{1}{2} \bar{\psi}(X) \tau^a \gamma_{\mu} \psi(X), \label{eq. noncons} \\
 \tilde{V}_{\mu}^a(X) &= \frac{1}{4} \Big( \bar{\psi}(X) \big( \gamma_{\mu}-1\big)U_{\mu}(X) \tau^a
\psi(X+\mu) + \nonumber \\ &+ \bar{\psi}(X+\mu) \big( \gamma_{\mu}+1 \big) U^{\dagger}_{\mu}(X)
\tau^a \psi(X) \Big), \label{eq. cons}
\end{align}
but differ in that $\tilde{V}_{\mu}^a$ is conserved by the lattice version of symmetry
transformations Eq.~\eqref{eq. symmetry}, whereas $V_{\mu}^a$ is not. We can relate these two
currents by a finite normalization factor denoted by $Z_{V}^{\MSb}$ defined as
\begin{equation}
% Z_{\tilde{V}}^{\MSb} \tilde{V}^a_{\mu} \equiv 
\tilde{V}^a_{\mu} = Z_{V}^{\MSb} V^a_{\mu}, 
\end{equation}
where we choose the $\MSb$ scheme only for definiteness. 
In the X-scheme, one has a similar condition  
\begin{equation}
 Z_{\tilde{V}}^{X} \tilde{V}^a_{\mu} = Z_{V}^{X} V^a_{\mu}. 
\end{equation}
In this case, however, even the conserved current receives 
a renormalization since the renormalization 
condition Eq.~\eqref{eq. condition} is incompatible with the symmetry given by Eq.~\eqref{eq.
symmetry} and thus the non-renormalization theorem does not hold. Note that this fact
already appears in continuum perturbation theory and is hence
no lattice specific effect. 
However, since the two discretizations given by Eqs.~\eqref{eq. noncons} and \eqref{eq. cons} agree
in the continuum limit, the
conversion to the $\MSb$ scheme through a perturbative conversion factor is the same for
$Z^X_{\tilde{V}} \tilde{V}_{\mu}$ and $Z^X_{V} V_{\mu}$.

\subsection{Perturbation theory results}
\label{sec. perturbative}

\subsubsection{Correlation functions}
At weak coupling, the functional form of the continuum counterparts  of the correlation functions
$C_{\Gamma \Gamma}(X)$ introduced by Eq.~\eqref{eq. correlation functions}, which are denoted by
$\Pi_{\Gamma}$ in the continuum theory, is known up to fourth order in the 
coupling constant \cite{Chetyrkin:2010dx}, namely
\begin{align}
 \Pi_P(X) = \Pi_S(X) &= \frac{3}{\pi^4 (X^2)^3} \Big( 1 + \sum_{n=1}^{\infty} \tilde{C}_n^S
\tilde{a}_s^n \Big), \\ 
 \Pi_{A,\mu \nu}(X) = \Pi_{V,\mu \nu}(X) &= \frac{6}{\pi^4 (X^2)^3} \left( \left( \frac{1}{2}
\delta_{\mu \nu}- \frac{X_{\mu} X_{\nu}}{X^2}\right) \tilde{C}^V + \delta_{\mu \nu} \tilde{D}^V
\right), 
\end{align}
with
\begin{equation}
 \tilde{C}^V = 1 + \sum_{n=1}^{\infty} \tilde{C}_n^V \tilde{a}_s^n, \quad \tilde{D}^V =
\sum_{n=0}^{\infty} \tilde{D}_n^V \tilde{a}_s^n,
 \label{eq. coefficients} 
\end{equation}
where the values of the perturbative coefficients $\tilde{C}_n^S$, $\tilde{D}_n^V$ and
$\tilde{C}_n^V$ for $n \le 4$ are given in Ref.~\cite{Chetyrkin:2010dx}.
The equalities of the pseudoscalar and scalar, as well as the vector and axial correlation
functions, follow from the assumption of quarks being massless and of working in the flavour-charged
sector in which the disconnected graphs are absent.
In the above expressions:
\begin{equation}
 \tilde{a}_s = \frac{\tilde{\alpha}_s(\tilde\mu=1/X)}{\pi},
\end{equation}
where
$\tilde{a}_s$ is defined in the $\MSt$ scheme \cite{Chetyrkin:2010dx} which is related to the $\MSb$
scheme by a shift of the renormalization scale: 
\begin{equation}
 \bar\mu\equiv \mu_{\MSb} = 2 e^{-\gamma_E} \mu_{\MSt} \approx 1.12 \mu_{\MSt}\equiv1.12\tilde{\mu}.
\end{equation}
The relation between $\MSt$ quantities and their $\MSb$ counterparts is
\begin{equation}
 \Pi^{\MSt}(X;\,\tilde\mu=1/X) = \Pi^{\MSb}(X;\, \bar\mu=2 e^{-\gamma_E} \tilde\mu), 
\end{equation}
and
\begin{equation}
 \tilde{a}_s(\tilde\mu) = a_s(\bar\mu=2 e^{-\gamma_E} \tilde\mu).
\end{equation}
$a_s$ is defined as $\alpha_s/\pi = g^2/(2 \pi)^2 $ with
\begin{equation}
 \mu^2 \frac{d a_s }{d\mu^2} = a_s \beta(a_s) \equiv - \sum_{i\ge 0}\beta_i a_s^{i+2}
\end{equation}
and the beta function in the $\MSb$ scheme is defined by
\begin{equation}
 \beta(a_s) = - \sum_{n=0}^{\infty} \beta_n a_s^{n+1},
\end{equation}
with the 4-loop $\beta_n$ coefficients computed in Ref.~\cite{vanRitbergen:1997va}.

\subsubsection{Perturbative conversion from the X-space scheme to the $\MSb$-scheme}
\label{sec. conversion}

The natural scale for the transition between the $\MSt$ scheme and the X-space scheme is
\begin{equation}
 \tilde{\mu}_0^2 = \frac{1}{X_0^2}.
\end{equation}
This is equivalent to the transition between the $\MSb$ scheme and the X-space scheme at a scale
\begin{equation}
\bar\mu^2_0 = \frac{4}{X_0^2} e^{-2\gamma_E}. 
\end{equation} 
For the correlation functions, we can write
\begin{equation}
 \Pi^X(X_0) = \Big( \frac{Z^X(X_0)}{Z^{\MSt}(\tilde{\mu}_0)} \Big)^2 \tilde{\Pi}(\tilde{\mu}_0) =
 \Big( \frac{Z^X(X_0)}{Z^{\MSb}(\bar\mu_0)} \Big)^2 \Pi(\bar\mu_0),  
\end{equation}
which yields the desired ratios needed for the conversion between the different schemes:
\begin{equation}
 \frac{Z^{\MSb}(\bar\mu_0)}{Z^X(X_0)} = \frac{Z^{\MSt}(\tilde{\mu}_0)}{Z^X(X_0)} =
\sqrt{\frac{\Pi^{\MSt}(\tilde{\mu}_0)}{\Pi(X_0)\big|_{\free}}} 
\end{equation}
In Ref.~\cite{Chetyrkin:2010dx}, these ratios have been computed perturbatively:
\begin{equation}
\label{eq:convS}
 \frac{Z^{\MSt}_S(\tilde{\mu}_0)}{Z_S^X(X_0)} = 1 + \sum_{n=1}^{\infty} \delta_n^S
\tilde{a}_s(\tilde{\mu}_0)^n
\end{equation}
and 
\begin{equation}
\label{eq:convV}
 \frac{Z^{\MSt}_V(\tilde{\mu}_0)}{Z_V^X(X_0)} =  1 + \sum_{n=1}^{\infty} \delta_n^V
\tilde{a}_s(\tilde{\mu}_0)^n
\end{equation}
with known numerical values of the coefficients $\delta_n^S$ and $\delta_n^V$ for $n \le 4$.

\subsubsection{Perturbative evolution of the renormalization constants}
\label{sec. evolution}

Given a renormalization constant at a given scale $\mu$ we can obtain perturbatively its value
at another scale $\mu'$ through the renormalization group equation
(staying solely in the $\MSb$-scheme)
\begin{equation}
\label{eq:evol}
 Z_{\Gamma}^{\MSb}(\mu') = \frac{c^{\MSb}_{\Gamma}(\mu')}{c^{\MSb}_{\Gamma}(\mu)}
Z_{\Gamma}^{\MSb}(\mu),
\end{equation}
where the perturbative expansion of the function $c^{\MSb}_{\Gamma}(\mu)$ for $\Gamma = \{S, P\}$,
related to the quark mass anomalous dimension, was computed in Refs.~\cite{Chetyrkin:1997dh},
\cite{Vermaseren:1997fq}.
For $\Gamma = \{V, A\}$, the Ward identities for the axial and vector
currents imply that their renormalization constants in the $\MSb$ scheme
are scale independent, which
is equivalent to saying that their anomalous dimensions vanish identically.

\section{Analysis strategy and example}
\label{sec. analysis}

\subsection{Lattice setup}
\label{sec. setup}
We apply the X-space method of extracting RCs for $N_f=2$ dynamical
ensembles, generated by ETMC \cite{Boucaud:2007uk}, \cite{Boucaud:2008xu}, \cite{Baron:2009wt},
using the tree-level Symanzik improved gauge action and 2 flavours of twisted mass
fermions.

We consider four values of the inverse gauge coupling $\beta$, corresponding to lattice spacing
values of between $\approx$ 0.04 and 0.08 fm.
The parameters of our ensembles are summarized in Tab. \ref{tab:ensembles}.

\begin{table}[t!]
\begin{center}
\begin{tabular}[b]{ccccccc}
\hline
\hline 
$\beta$ & $\kappa$ & $L^3\times T$ & $a\mu$ & Label &  $a$ $[\textrm{fm}]$ & $L$
$[\textrm{fm}]$ \\%& $r_0/a$\\
\hline
3.9 & 0.160856 & $24^3\times48$ & 0.0040 & B40.24 & 0.0790(26) &  1.9 \\% & 5.35(4)\\
 &  &  & 0.0064 & B64.24 \\
 &  &  & 0.0085 & B85.24 \\
 &  &  & 0.0150 & B150.24 \\
\hline
4.05 & 0.157010 & $32^3\times64$ & 0.0030 & C30.32 & 0.0630(20) & 2.0 \\%& 6.71(4)\\
& & & 0.0060 & C60.32\\
& & & 0.0080 & C80.32\\
\hline
4.20 & 0.154073 & $48^3\times96$ & 0.0020 & D20.48 & 0.05142(83) & 2.4 \\%& 8.36(6)\\
\hline
4.35 & 0.151740 & $32^3\times64$ & 0.00175 & E17.32 & 0.0420(17) & 1.3 \\%& 9.82(4)\\
\hline
\hline
\end{tabular}
\end{center}
\caption{\label{tab:ensembles} Ensembles and their parameters used for the calculations. 
Results for the lattice spacing are taken from Ref.~\cite{Baron:2009wt} (ensembles B,C,D) and
Ref.~\cite{Jansen:2011vv} (ensemble E).}
\end{table} 

We perform our investigation for four types of correlation functions:
\begin{align}
 C_{SS}(X) = \langle S(X) S(0) \rangle, & \quad C_{PP}(X) = \langle P(X) P(0) \rangle, \nonumber\\
 C_{VV}(X) = \sum_{\mu} \langle V_{\mu}(X) V_{\mu}(0) \rangle, & \quad C_{AA}(X) = \sum_{\mu}
\langle A_{\mu}(X) A_{\mu}(0) \rangle.\nonumber, 
\end{align}
which allows us to extract the four renormalization constants $Z_S$, $Z_P$, $Z_A$, $Z_V$, as well
as the ratios $Z_P/Z_S$ and $Z_V/Z_A$.

\subsection{Outline of analysis procedure}
\label{sec. procedure}
Our analysis procedure for all ensembles is shortly outlined below. In the next subsection, we
provide the details for each step and illustrate it with an example of the extraction of $Z_P$ for
ensemble E17.32, see table~\ref{tab:ensembles}:
\begin{enumerate}
\item Choose ensemble and the correlation function of interest.
\item For each configuration, average over sites that are equivalent with respect to the hypercubic
symmetry (taking the anisotropy of the lattice into account).
\item For each point $X\equiv(x,y,z,t)\equiv(x_1,x_2,x_3,x_4)$, take the ensemble average.
\item Correct for tree-level discretization effects.
\item Apply some cuts to eliminate points with large remaining
discretization effects. This amounts to choosing a direction in coordinate space,
defined by an angle $\theta(X)$ between the position vector of point $X$ and the direction
(1,1,1,1). The cut consists in selecting only points with $\theta(X)<\theta_{\rm max}$,
where $\theta_{\rm max}$ is chosen empirically. 
\item Average over \emph{neighbouring} directions with equal $X^2=\sum_\mu x_\mu x_\mu$.
\item Compute the $X_0^2$ dependence of RCs in the X-space scheme
$Z^X_{\Gamma}(X_0;a,\mu)$, applying the renormalization condition \eqref{eq. condition} at
different renormalization scales $X_0$.
\item Choose the interval to extract the final value of RC: $a\ll X_0 \ll
\Lambda^{-1}_{\mathrm{QCD}}$.
\item (comparison with other methods) Convert to another renormalization scheme of interest, e.g.
the $\MSb$ scheme, evolve to the reference scale of interest, e.g. $\mu=2$ GeV.
\end{enumerate}

\subsection{Detailed example -- $Z_P$}
We now give a detailed example of the complete analysis procedure drafted in the previous
subsection. 
For the convenience of the Reader, we explicitly mark the stage of the analysis procedure a given
part of the text corresponds to. 

\textbf{Stage 1.} We demonstrate the extraction of the renormalization constant $Z_P$ for the ensemble
E17.32.

\textbf{Stage 2.} The hypercubic symmetry implies that the correlation function computed at points
which are equivalent from the point of view of this symmetry should take the same value. This is
indeed true in the free theory, but in the interacting case this symmetry is broken in the
process of stochastic generation of gauge field configurations by the Hybrid Monte Carlo method,
resulting in different values of the correlation function for points equivalent with respect to the
hypercubic symmetry. For our analysis, such points should be averaged over.
Given some point $(x,y,z,t)$, the equivalent points are the 6 permutations: $(x,y,z,t)$,
$(y,z,x,t)$, $(z,x,y,t)$, $(x,z,y,t)$,
$(y,x,z,t)$, $(z,y,x,t)$. Note that in order to take into account lattice anisotropy ($T=2L$), we
treat $t$ differently -- indeed even in the free theory the value of the correlation function
computed at $(x,y,z,t)$ and e.g. $(t,x,y,z)$ is different (unless $z=t$). In addition to these six
permutations, the equivalent points include six permutations of all of the following points:
$(L-x,y,z,t)$, $(x,L-y,z,t)$, $(x,y,L-z,t)$, $(L-x,L-y,z,t)$, $(L-x,y,L-z,t)$, $(x,L-y,L-z,t)$,
$(L-x,L-y,L-z,t)$. Finally, in all of the above points one can replace $t\rightarrow T-t$. This
gives in general 96 equivalent points, although this number may be reduced for points with some of
the coordinates zero or equal to each other.

\begin{figure}
\begin{center}
\includegraphics
[width=0.34\textwidth,angle=270]
{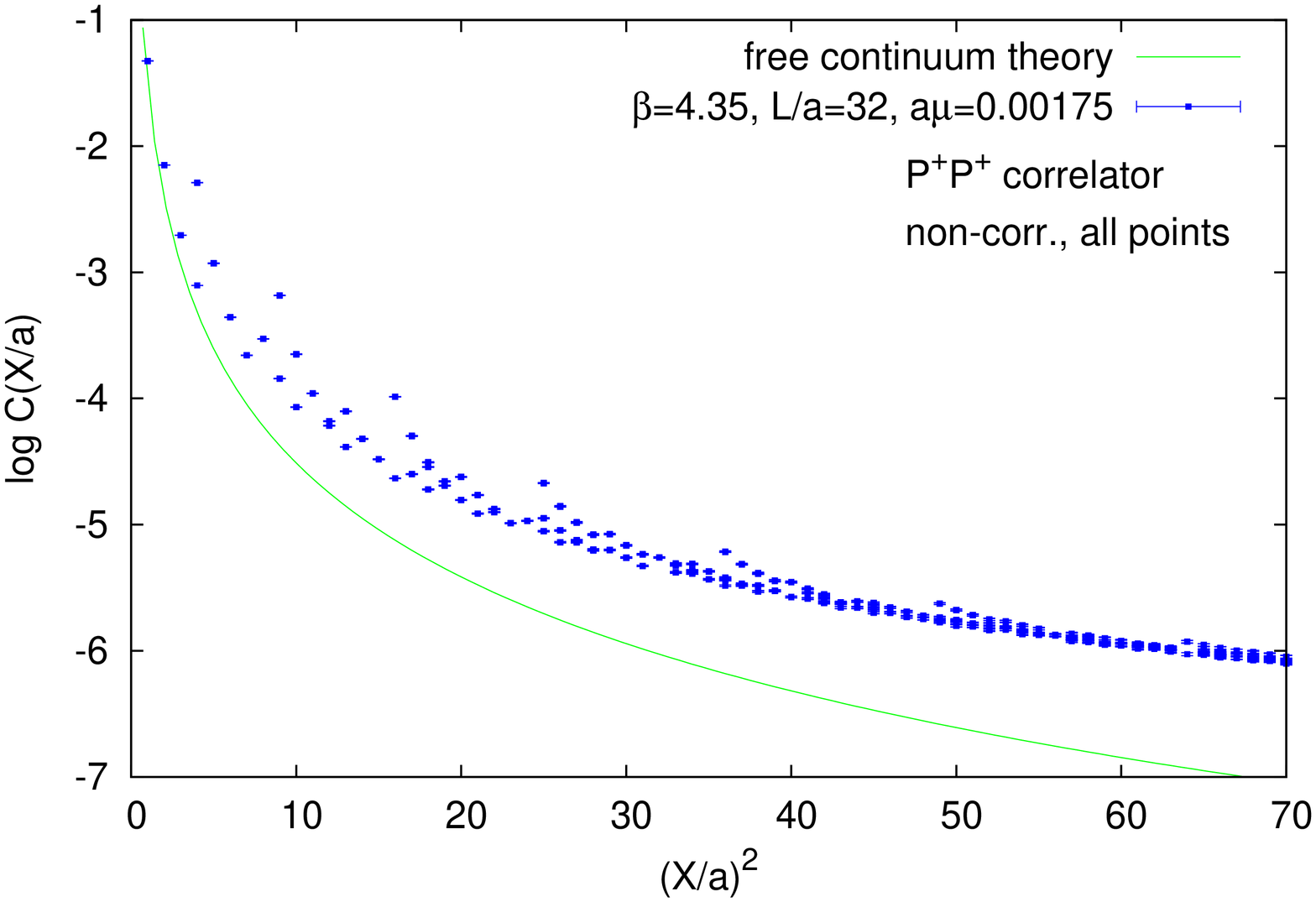}
\includegraphics
[width=0.34\textwidth,angle=270]
{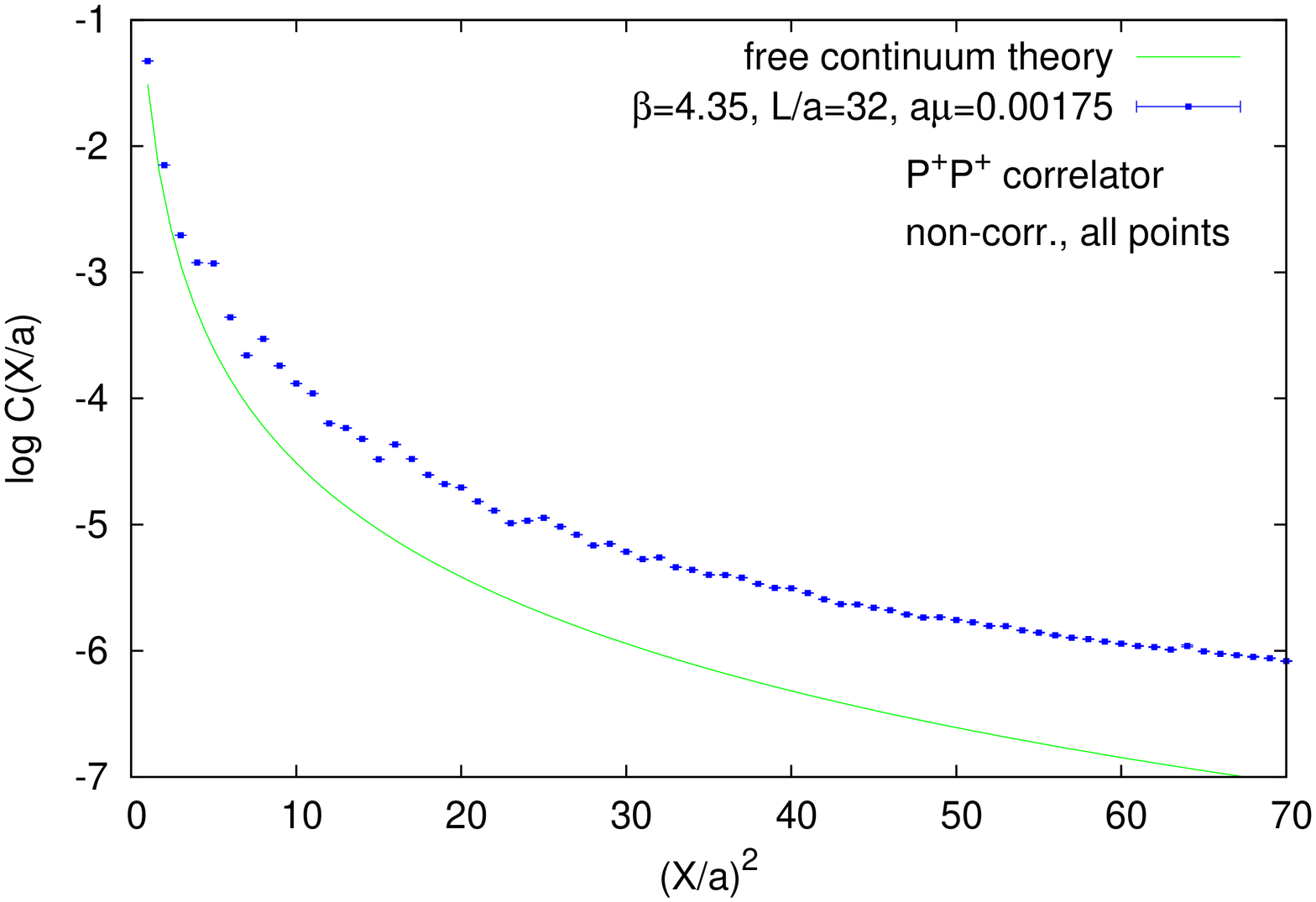}
\end{center}
\caption{\label{fig:corr_avg} The PP correlation function for ensemble E17.32, (left) plotted
vs. $X^2$ averaged over equivalent point obtained by permutations (Stage 2), 
(right) averaged over all points with the same $X^2$.}
\end{figure}

\textbf{Stage 3.} The above averaging is performed for every gauge field configuration separately.
As a result, the correlation functions are fully symmetric with respect to the hypercubic symmetry,
as in the free theory case. We then compute the ensemble average of the correlators averaged over
equivalent sites, together with the corresponding statistical errors. The resulting values of the PP
correlator (ensemble E17.32) are plotted in Fig.~\ref{fig:corr_avg}(left) as a function of $X^2$.
Even after exploiting the permutation symmetry, there still remain points
with the same value of $X^2$, e.g. (2,2,2,2) and (4,0,0,0), leading to 
additional values of the correlation functions for the same $X^2$. 
A possible treatment would be to average over such points 
and the result of such averaging is shown in
Fig.~\ref{fig:corr_avg}(right). The obtained correlator shows very irregular behaviour, especially
for short distances that are crucial from the viewpoint of extracting RCs and making connection to
perturbation theory.

\begin{figure}
\begin{center}
\includegraphics
[width=0.34\textwidth,angle=270]
{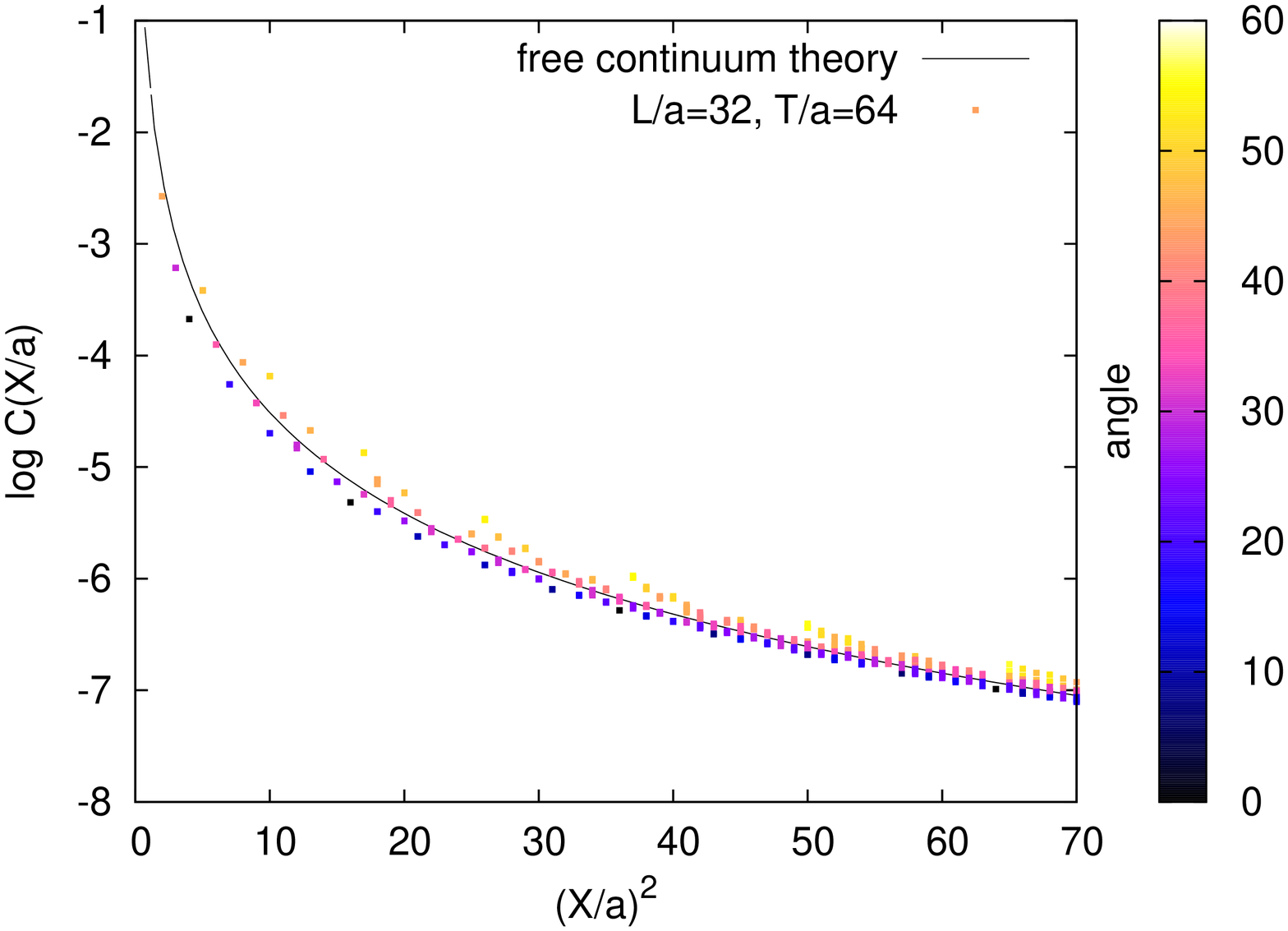}
\includegraphics
[width=0.34\textwidth,angle=270]
{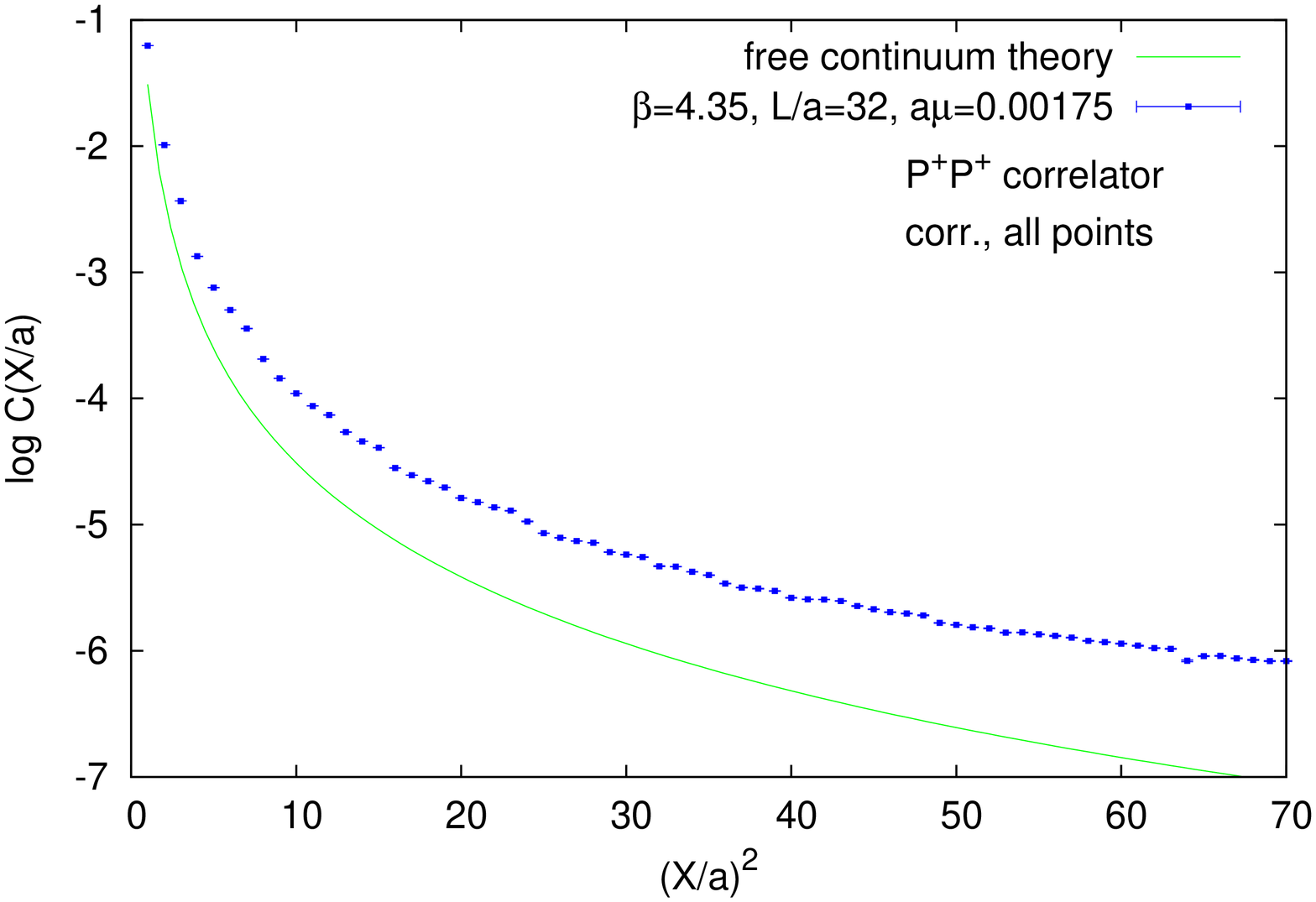}
\end{center}
\caption{\label{fig:corr_avg_corrected} (left) The PP correlation function in the free-field lattice
theory, $L/a=32$, $T/a=64$, $a\mu=0$. The colour scale shows the angle of the position vectors of
all points with respect to the direction (1,1,1,1) of the hypercube. (right) The 
(lattice) tree-level corrected PP
correlation function for ensemble E17.32, averaged over points with the same $X^2$.}
\end{figure}

\textbf{Stage 4.} Clearly, the non-smoothness of the correlator at this stage comes from
discretization effects. Some part of these effects appears already in the free-field theory. This is
illustrated in the left panel of Fig.~\ref{fig:corr_avg_corrected}. Comparing it to
Fig.~\ref{fig:corr_avg}(left), one can notice the same kinds of structures (``tails'') in the free
and interacting correlators. This suggests that it is possible to reduce the cut-off effects
in the interacting theory correlator by subtracting tree-level discretization effects. Following
Ref.~\cite{Gimenez:2004me}, we computed for each correlator function type $C_{\Gamma\Gamma}$ a
correction factor $\Delta_{\Gamma}$, defined as the ratio of the free correlator on the lattice over
the continuum one calculated in infinite volume and in the chiral limit:
\begin{equation}
\label{eq:delta}
 \Delta_{\Gamma} = \frac{\langle \oo_{\Gamma}(X) \oo_{\Gamma}(0)
\rangle_{\lat}^{\free}}{\langle \oo_{\Gamma}(X) \oo_{\Gamma}(0)
\rangle_{\cont}^{\free}}=
\frac{\langle \oo_{\Gamma}(X) \oo_{\Gamma}(0)
\rangle_{\lat}^{\free}}{\frac{c}{\pi^4(X^2)^3}},
\end{equation}
where $c=3$ for PP and SS correlators, $c=6$ for VV and AA
correlators.
Thus, we obtained the corrected correlation functions, $C'_{\Gamma
\Gamma}(X)$:
\begin{equation}
\label{eq:cprim}
 C'_{\Gamma \Gamma}(X) = \frac{C_{\Gamma \Gamma}(X)}{\Delta_{\Gamma}(X)}.
\end{equation}

As an effect of applying this correction to the PP correlator, we obtain much reduced spread in
the values of correlators corresponding to the same value of $X^2$. We also show the effect of the
free theory correction for the correlator averaged over points with the same $X^2$
(Fig.~\ref{fig:corr_avg_corrected}(right)) -- it is much smoother than the one before the
free-field correction (Fig.~\ref{fig:corr_avg}(right)). However, the behaviour is still not smooth
enough to reliably extract renormalization constants (note the logarithmic scale for the correlator
values). This results, naturally, from the fact that the subtracted cut-off effects were only the
non-interacting theory ones. We expect that the remaining cut-off effects, 
$\Oag$ at leading order in lattice perturbation theory,
can cause substantial cut-off effects for different types of points (different directions defined by the
position vectors) -- hence averaging over points with the same $X^2$ is at this stage unjustified.
The role of $\Oag$ effects and their potential subtraction will be discussed
further below.

So far, our analysis procedure closely followed Ref.~\cite{Gimenez:2004me}. The quality of our
averaged correlators, after the tree-level correction, is similar, but somewhat worse, than the one
of Ref.~\cite{Gimenez:2004me}. 
Clearly, further reduction of the cut-off effects that remain in our averaged correlators is 
desirable.

\textbf{Stage 5.} For this purpose, we employ an analogue of an idea widely used in momentum space
computations
\cite{Leinweber:1998uu}, \cite{Becirevic:1999uc}, \cite{Becirevic:1999hj}, \cite{deSoto:2007ht}.
Here we summarize its essence.
On the lattice, Green functions depend on lattice momenta
$\hat p_\mu\equiv\frac{2}{a}\sin\left(\frac{ap_\mu}{2}\right)$. Developing $\hat
p^2\equiv\sum_\mu p_\mu^2$ in terms of $a$ gives:
\begin{equation}
\hat p^2=p^2-\frac{a^2}{12}p^{[4]}+\frac{a^4}{360}p^{[6]}-\frac{a^6}{20160}p^{[8]}+\ldots,
\end{equation}
where the invariants $p^{[n]}\equiv\sum_\mu p_\mu^n$.
Hence, for momenta with the same $p^2$, larger artefacts are expected for the ones with larger
$p^{[4]}$.
The underlying reason is the reduction of continuum isometry group $O(4)$ to the discrete isometry
group $H(4)$ \footnote{For this discussion, we ignore the fact that for anisotropic lattices, the
group $H(4)$ is broken to $H(3)$, which has minor practical implications for our case of
interest.}. The orbits of $O(4)$ are labeled by a single invariant $p^2$, while lattice
momenta which belong to the same orbit of $O(4)$ do not have to belong to the same orbit of $H(4)$.
Already for $p^2=4$, a single $O(4)$ orbit splits into two $H(4)$ orbits, corresponding to momenta
(1,1,1,1) ($p^{[4]}=4$) and (2,0,0,0) ($p^{[4]}=16$). 
The $H(4)$ orbits labeled by the same $p^2$ and $p^{[4]}$ can further split into orbits with
different $p^{[6]}$ etc. (e.g. momenta (0,7,7,0) and (2,5,8,0) have the same $p^2=98$ and
$p^{[4]}=4802$, but $p^{[6]}=235298$ for the former and $278498$ for the latter).

In momentum space, the above argument implies that in order to have minimal cut-off effects, one
should consider momenta with a possibly ``democratic'' distribution of the total momentum $p^2$ over
different points, since this minimizes the value of the $p^{[4]}$ invariant. By analogy, one can
expect that in position space the points least affected by discretization effects are also the
``democratic'' ones, i.e. the ones close to the hypercubic diagonal (1,1,1,1).
These are the points with the smallest values of $X^{[4]}\equiv\sum_\mu x_\mu^4$ among
the ones with a given $X^2$. 

That this expectation is fulfilled, one can again observe in
Fig.~\ref{fig:corr_avg_corrected}(left). For each point, one can define an angle between the
position vector of point $X$ and the direction (1,1,1,1) -- let us denote it $\theta(X)$.
Different values of this angle correspond in Fig.~\ref{fig:corr_avg_corrected} to
different colours. We observe a very regular behaviour -- the points that make up the ``tails'' are
the ones with large values of $\theta(X)$ and the larger this value, the further away from the
continuum curve a given point lies. We may suspect that such large $\theta(X)$ points 
enhance cut-off effects both in the free theory (as shown in the plot) and in the interacting
theory. In particular, for such points the tree-level correction \eqref{eq:cprim} is not enough.
Therefore, we chose to avoid them, by defining the following criterion: keep only points with
$\theta(X)\leq\theta_{\rm max}$.
Naively, the choice of $\theta_{\rm max}=0$ should be the best one, as it guarantees the smallest
value of $X^{[4]}$ among all points with a given $X^2$. 
However, such choice is obviously too restrictive, as it would keep only very few points, none of
them or at most one satisfying the restriction $a\ll X_0 \ll \Lambda^{-1}_{\mathrm{QCD}}$.
Moreover, Fig.~\ref{fig:corr_avg_corrected}(left) shows that the analogy with momentum space is not
exact -- the most ``democratic'' points lie systematically below the free continuum curve and the
``optimal'' value of $\theta(X)$ is around 20-30 degrees, i.e. such values of $\theta(X)$ lead to
free lattice correlator values nearest to the free continuum values.
Since the free lattice computation is performed at finite $L/a$, the natural question to ask at
this point is whether this apparently counterintuitive behaviour is related to finite size effects.
Simulating on larger lattices, we eliminated this possibility -- finite size effects are negligible
if the coordinates $x,y,z,t$ are small in relation to $L$ and $T$. Since we are interested in the
small distance behaviour, this condition is satisfied. 

In the interacting theory, the situation is, of course, even more complex. It is again true that
the points constituting the ``tails'' are the ones with largest values of $\theta(X)$. We clearly
see that these points are responsible for the non-smoothness of the correlator in
Fig.~\ref{fig:corr_avg_corrected}(right). Hence, we decided to choose $\theta_{\rm max}=30$
degrees. This value does not have a rigorous justification, but was found to give the best
improvement in the smoothness of our correlators while still keeping a 
sufficient number of points for a reliable extraction of the RCs.

\textbf{Stage 6.} What is more, for most values of $X^2$ of
interest to us, the criterion $\theta(X)<\theta_{\rm max}$ selects only points being permutations
of $(x,y,z,t)$ interchanging spatial and
temporal indices (e.g. $(y,z,t,x)$). Such points are not strictly equivalent with respect to the
hypercubic symmetry, because of the anisotropy of our lattices in the time direction -- hence, we
did not average over them in the initial stage of our procedure. Thus, it allowed us to apply a
slightly different tree-level correction for such \emph{almost} equivalent points. After this
correction, the correlator values for such cases are compatible within statistical error and can be
safely averaged over. 
With the choice $\theta_{\rm max}=30$, we do not have to average points that correspond to
different types of points, like e.g. (2,2,2,2) and (4,0,0,0), since our ``democratic'' cut
eliminates such necessity. In this way, we don't mix points with possibly very different cut-off
effects (such points are then non-compatible within the statistical error even after the
free-field correction).

\begin{figure}
\begin{center}
\includegraphics
[width=0.34\textwidth,angle=270]
{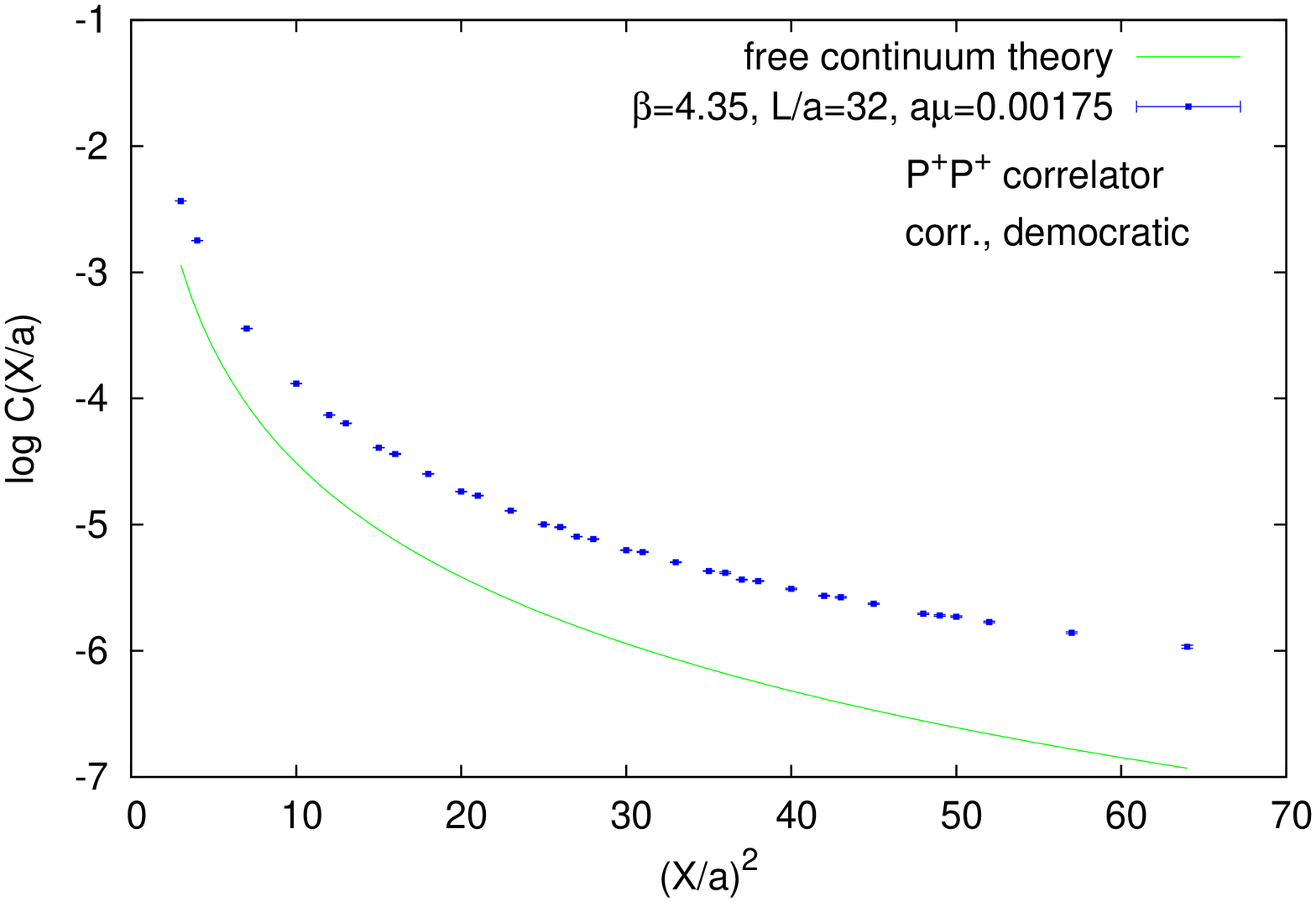}
\includegraphics
[width=0.34\textwidth,angle=270]
{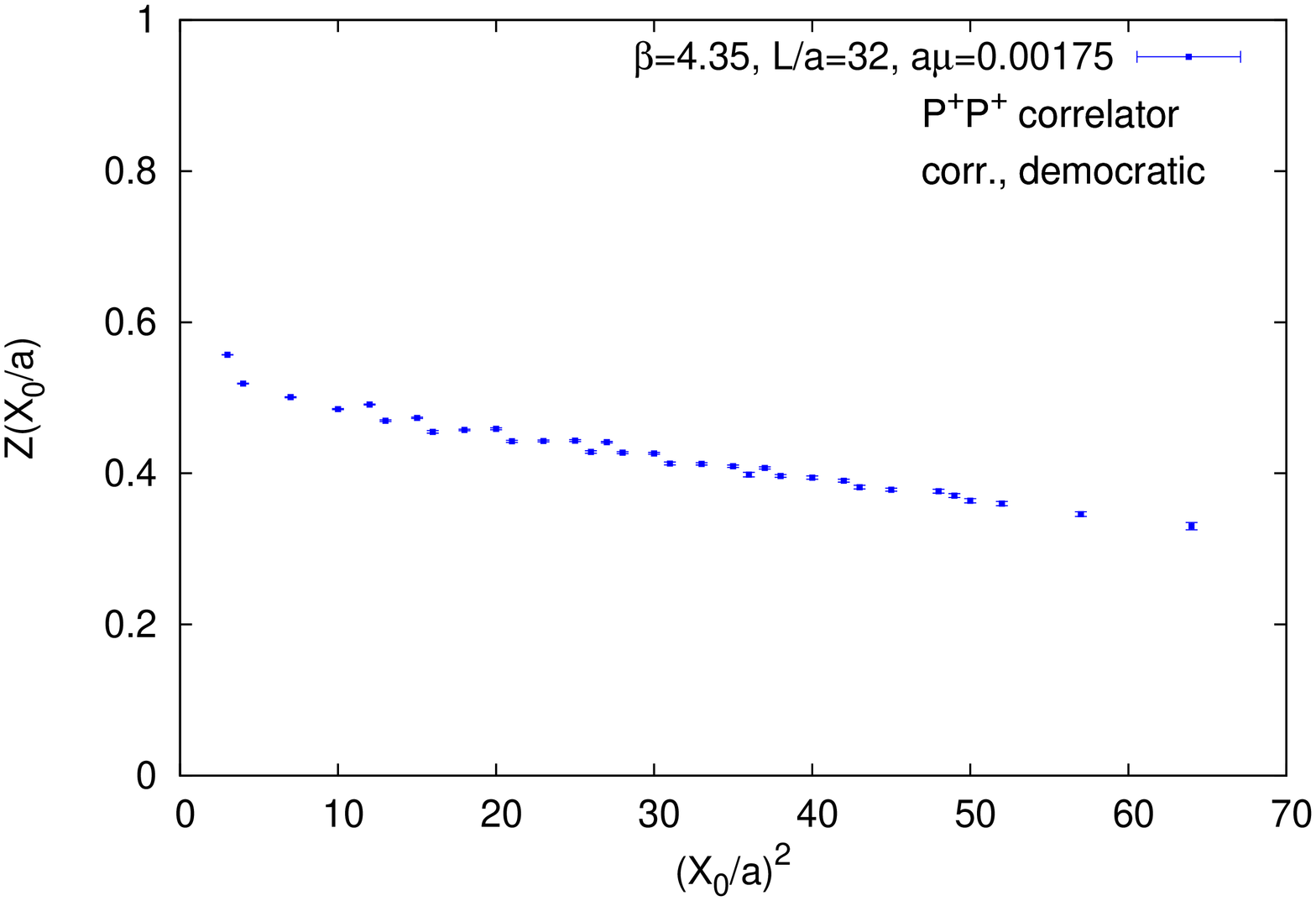}
\end{center}
\caption{\label{fig:corr_avg_democratic} (left) The corrected PP
correlation function for ensemble E17.32. Only ``democratic'' points were used ($\theta(X)\leq30$
degrees). Averaging over points with the same $X^2$ involves only
points interchanging spatial and
temporal indices. (right) The corresponding values of the renormalization constant $Z_P$ vs. the
renormalization scale $X_0^2$.}
\end{figure}

The corrected PP correlator after the ``democratic'' cut is shown in
Fig.~\ref{fig:corr_avg_democratic}(left). In accordance with our expectations, the correlator curve
is much smoother than the one after only free-field correction. Obviously, we have not eliminated
all cut-off effects and therefore some non-smoothness is still present.

\textbf{Stage 7.} Now, in order to extract any renormalization constant $Z_\Gamma$ (at some
renormalization scale $X_0$), we impose the X-space renormalization condition directly on the
corrected correlation function (using solely ``democratic'' points), i.e.:
\begin{equation}
\frac{\langle \oo^X_{\Gamma}(X) \oo^X_{\Gamma}(0) \rangle'}{\langle
\oo_{\Gamma}(X) \oo_{\Gamma}(0) \rangle_{\cont}^{\free}}\Big|_{X^2=X_0^2} =
\frac{\langle \oo^X_{\Gamma}(X) \oo^X_{\Gamma}(0) \rangle}{\langle
\oo_{\Gamma}(X) \oo_{\Gamma}(0) \rangle_{\lat}^{\free}}\Big|_{X^2=X_0^2} = 1.
\end{equation}
Since $\oo^X_{\Gamma}(X) = Z^X_{\Gamma}(X_0) \oo_{\Gamma}(X)$,
we have:
\begin{equation}
 Z^X_{\Gamma}(X_0) = \sqrt{\frac{C_{\Gamma
\Gamma}(X_0)_{\cont}^{\free}}{C'_{\Gamma \Gamma}(X_0)}}
= \sqrt{\frac{C_{\Gamma
\Gamma}(X_0)_{\lat}^{\free}}{C_{\Gamma \Gamma}(X_0)}}.
\end{equation} 

Using every value of $X^2$ for which the corrected ``democratic'' correlators are available, we
vary the renormalization scale $X_0^2$ and hence obtain the $X_0^2$ dependence of our
renormalization
constants in the X-space scheme. An example for $Z_P^X$ is plotted in
Fig.~\ref{fig:corr_avg_democratic}(right). It shows that $Z_P$ is scale-dependent (as expected) and
reveals that indeed there is still some non-smoothness in the data (concealed in the correlator
plot, because of the logarithmic scale), attributed to remaining cut-off effects,
$\Oag$ at the leading order.

\textbf{Stages 8, 9.} In principle, the computed values of the X-space renormalization constants
$Z_\Gamma^X$ can be used to renormalize the corresponding operators. In practice, one usually wants
operators renormalized in the $\MSb$ scheme, at some reference scale, often chosen to be $\mu=2$
GeV. Therefore, we convert our X-space renormalization constants to the $\MSb$ scheme, using 4-loop
conversion formulae, as described in Sec.~\ref{sec. conversion}.

For renormalization constants that are scale-independent in the $\MSb$ scheme ($Z_A$, $Z_V$), the
final values in this scheme are obtained in the following way:
\begin{equation}
Z_{V,A}^X(1/X_0)\xrightarrow[{\rm Eq.}\,\tiny\eqref{eq:convV}]{\mathrm{conversion}}
Z_{V,A}^{\MSt}(\mu=1/X_0) = Z_{V,A}^{\MSb}(\mu=2e^{-\gamma_E}/X_0) = 
Z_{V,A}^{\MSb}(2\,\mathrm{GeV}). 
\end{equation} 
The value in the X-space scheme at the scale $\mu=1/X_0$ is converted to the intermediate scheme
$\MSt$ at the same scale. Since the value in the $\MSt$ scheme is equal to the value in the $\MSb$
scheme at the scale $\mu=2e^{-\gamma_E}/X_0$ and is moreover scale-independent, the
perturbative conversion is the only step needed to obtain $Z_{V,A}^{\MSb}(2\,\mathrm{GeV})$.
Note also that the scale independence of $Z_{V,A}$ in the $\MSb$ scheme and the scale dependence of
the conversion factor (rather mild, if the renormalization scale $X_0$ is not too small) imply that
$Z_{V,A}$ computed in the X-space scheme should show an $X_0$-dependence. This is a direct
result of the breaking of chiral Ward identities.
In practice, it turns out that this $X_0$-dependence is rather mild, as we show in this 
work and was also observed in Ref.~\cite{Gimenez:2004me}.

In the case of scale-dependent renormalization constants ($Z_P$, $Z_S$) the procedure involves
additionally the evolution to the reference scale:
\begin{align}
Z_{P,S}^X(1/X_0)\xrightarrow[{\rm Eq.}\,\tiny\eqref{eq:convS}]{\mathrm{conversion}}
Z_{P,S}^{\MSt}(\mu=1/X_0) =
Z_{P,S}^{\MSb}(\mu=2e^{-\gamma_E}/X_0) \nonumber\\
\xrightarrow[{\rm Eq.}\,\tiny\eqref{eq:evol}]{\mathrm{evolution}}
Z_{P,S}^{\MSb}(2\,\mathrm{GeV}).
\end{align} 
The value in the X-space scheme at the scale $\mu=1/X_0$ is converted to the $\MSt$ scheme at the
same scale, equal to the $\MSb$ scheme value at the scale $\mu=2e^{-\gamma_E}/X_0$. Using
renormalization group equations (Sec.~\ref{sec. evolution}), we finally obtain
$Z_{P,S}^{\MSb}(2\,\mathrm{GeV})$.

In order to obtain reliable values of the renormalization constants
$Z_{\Gamma}^{\MSb}$ at the scale of 2 GeV, the perturbative formulae have to be 
applied for scales that
satisfy the condition $1/ X_0 \gg \Lambda_{\mathrm{QCD}}$. Moreover, scales close to the lattice
cut-off have to be avoided. Together, this yields the condition $a\ll X_0 \ll
\Lambda^{-1}_{\mathrm{QCD}}$, which imposes applicability constraints on the method -- the
condition can be satisfied only on lattices with fine enough lattice spacing.

\begin{table}[t!]
\begin{center}
\begin{tabular}[b]{ccccc}
\hline
\hline 
\multirow{2}{*}{$\beta$} & $a^{-1}$ & \multicolumn{3}{c}{Extraction range}\\
& [MeV] &$(X_0/a)^2$  & $\tilde\mu=1/X_0$ [MeV] & $\bar\mu=2e^{-\gamma_E}/X_0$ [MeV]\\
\hline
3.9 & 2500 & $[7,\,13]$ & $[695,\,945]$ & $[775,\,1060]$\\
4.05 & 3130 & $[7,\,15]$ & $[810,\,1185]$ & $[905,\,1325]$\\
4.2 & 3870 & $[10,\,18]$ & $[910,\,1225]$ & $[1020,\,1370]$\\
4.35 & 4700 & $[10,\,21]$ & $[1025,\,1490]$ & $[1150,\,1670]$\\
\hline
\end{tabular}
\end{center}
\caption{\label{tab:extraction} Extraction windows for RCs, satisfying the constraint 
$a\ll X_0 \ll \Lambda^{-1}_{\mathrm{QCD}}$ or, equivalently
$1/a\gg \mu \gg \Lambda_{\mathrm{QCD}}$ with $\Lambda_{\mathrm{QCD}}$ typically
about 300 MeV. In practice, we take $X_0\gtrsim 3a$.
The values of RCs in the X-space scheme at
the scale $1/X_0$ are perturbatively converted to the $\MSt$ scheme at the scale denoted by
$\tilde\mu$ and are equal to the $\MSb$ scheme values at the scale denoted by $\bar\mu$.}
\end{table} 

For our ensembles of interest, we have chosen the RCs extraction windows listed in
Tab.~\ref{tab:extraction}. Our choices depend on the values of the lattice spacing corresponding to
each value of the inverse bare coupling $\beta$. 
Besides the physical conditions for which the window has been chosen, 
the inclusion of
several points allows for an important check of the scale dependence of RCs. The values in the
X-space scheme depend on the scale. However, after conversion to the $\MSb$ scheme and evolution to
2 GeV, all RCs should be independent of the initial scale up to statistical accuracy,
non-subtracted
$\Oag$ and higher order discretization effects. If this comes out to be true 
in the final analysis, it means that
the lattice simulation reproduces correctly the quark mass anomalous dimension ($Z_P$, $Z_S$) or the
scale independence of $Z_V$, $Z_A$. Moreover, it provides a
strong hint that the extraction window was chosen appropriately.
For these reasons, we decided to go rather low in $1/X_0$, especially for the case of $\beta=3.9$.
Whereas 775 MeV might seem dangerously close to $\Lambda_{\mathrm{QCD}}$, the inclusion of points
near this scale does not spoil the constant behaviour of $\MSb(2\,\mathrm{GeV})$
RCs converted to the $\MSb$-scheme and evolved from X-space scheme values at different scales
$1/X_0$ 
to the reference scale of $2\,\mathrm{GeV}$ in the $\MSb$-scheme, as we will 
further discuss below. 

\begin{table}[t!]
%\begin{footnotesize}
\begin{center}
\begin{tabular}[b]{cccccc}
\hline
\hline 
\multirow{2}{*}{$(X_0/a)^2$} & $1/X_0$ & \multirow{2}{*}{$Z_{P}^X(1/X_0)$} & $\bar\mu$
 & \multirow{2}{*}{$Z_{P}^{\MSb}(\bar\mu$)} &
\multirow{2}{*}{$Z_{P}^{\MSb}(2\,\mathrm{GeV})$}\\
& $[\mathrm{MeV}]$ & & $[\mathrm{MeV}]$\\
\hline
10 & 1490 & 0.4850(8) & 1670 &  0.4685(8)(12)(17) & 0.4921(9)(49)(2)\\
12 & 1360 & 0.4911(6) & 1520 &  0.4724(6)(13)(19) & 0.5101(7)(55)(7)\\
13 & 1300 & 0.4697(10) & 1460 & 0.4508(10)(13)(20) & 0.4932(11)(55)(11)\\
15 & 1210 & 0.4734(8) & 1360 & 0.4526(8)(14)(22) & 0.5075(9)(62)(20)\\
16 & 1170 & 0.4549(17) & 1320 & 0.4340(16)(14)(22) & 0.4924(18)(62)(24)\\
18 & 1110 & 0.4574(10) & 1240 & 0.4347(9)(15)(23) & 0.5047(11)(69)(35)\\
20 & 1050 & 0.4589(13) & 1180 & 0.4346(12)(16)(24) & 0.5156(14)(76)(47)\\
21 & 1025 & 0.4425(14) & 1150 & 0.4183(13)(16)(24) & 0.5017(16)(77)(51)\\
\hline
\hline
\end{tabular}
\end{center}
\caption{\label{tab:example} Example of the conversion of X-space RCs values to the $\MSb$ scheme
and their subsequent evolution to the reference scale of 2 GeV. The scale
$\bar\mu=2e^{-\gamma_E}/X_0$.
The error for $Z_P^X$ is statistical.
Errors for $Z_{P}^{\MSb}$ are: 1st -- statistical, 2nd -- resulting from the uncertainty in lattice
spacing, 3rd -- resulting from the uncertainty of $\Lambda_{\MSb,N_f=2}$. See text for more
details. The chosen ensemble is E17.32.}
%\end{footnotesize}
\end{table} 

In Tab.~\ref{tab:example}, we give an explicit example of the conversion/evolution procedure for
$Z_P$. The third and fifth columns show the scale dependence of $Z_P$ in the X-space scheme and in
the $\MSb$ scheme, whereas the last column shows that the values converted from different scales
$1/X_0$ lead to similar values of $Z_{P}^{\MSb}(2\,\mathrm{GeV})$, with some fluctuations around the
central value. 
Tab.~\ref{tab:example} also shows how various types of errors enter at the stage of conversion and
evolution. The first error is always statistical (for the X-space value it is the only error).
The second error (systematic) results from the fact that uncertainties in the value of the lattice
spacing (listed in Tab.~\ref{tab:ensembles}) lead to uncertainties in the values of the scale
$1/X_0$
in physical units, which is, in turn, needed to calculate the value of the strong coupling constant
that enters the perturbative conversion or evolution formulae.
The third error (systematic) originates from the fact that we do not precisely know the value of
the scale $\Lambda_{\rm QCD}$. To emphasize that our present case is the $N_f=2$
one and we work in the $\MSb$ scheme, we denote it by $\Lambda_{\MSb,N_f=2}$.
Its value has recently been computed by the ALPHA Collaboration -- 310(20) MeV
\cite{Fritzsch:2012wq} and by ETMC -- 315(30) MeV \cite{Jansen:2011vv}.

\begin{figure}[t!]
\begin{center}
\includegraphics
[width=0.34\textwidth,angle=270]
{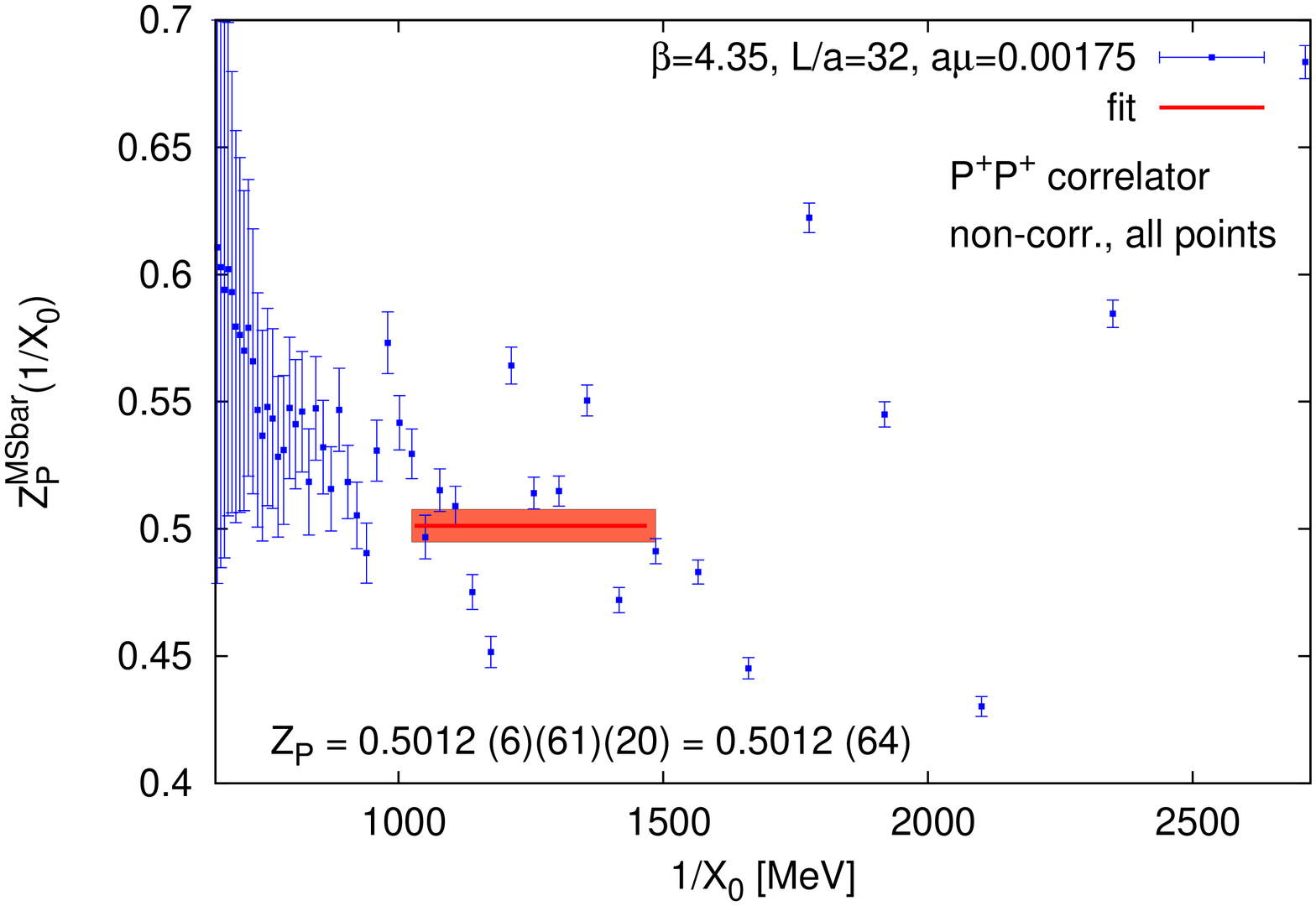}
\includegraphics
[width=0.34\textwidth,angle=270]
{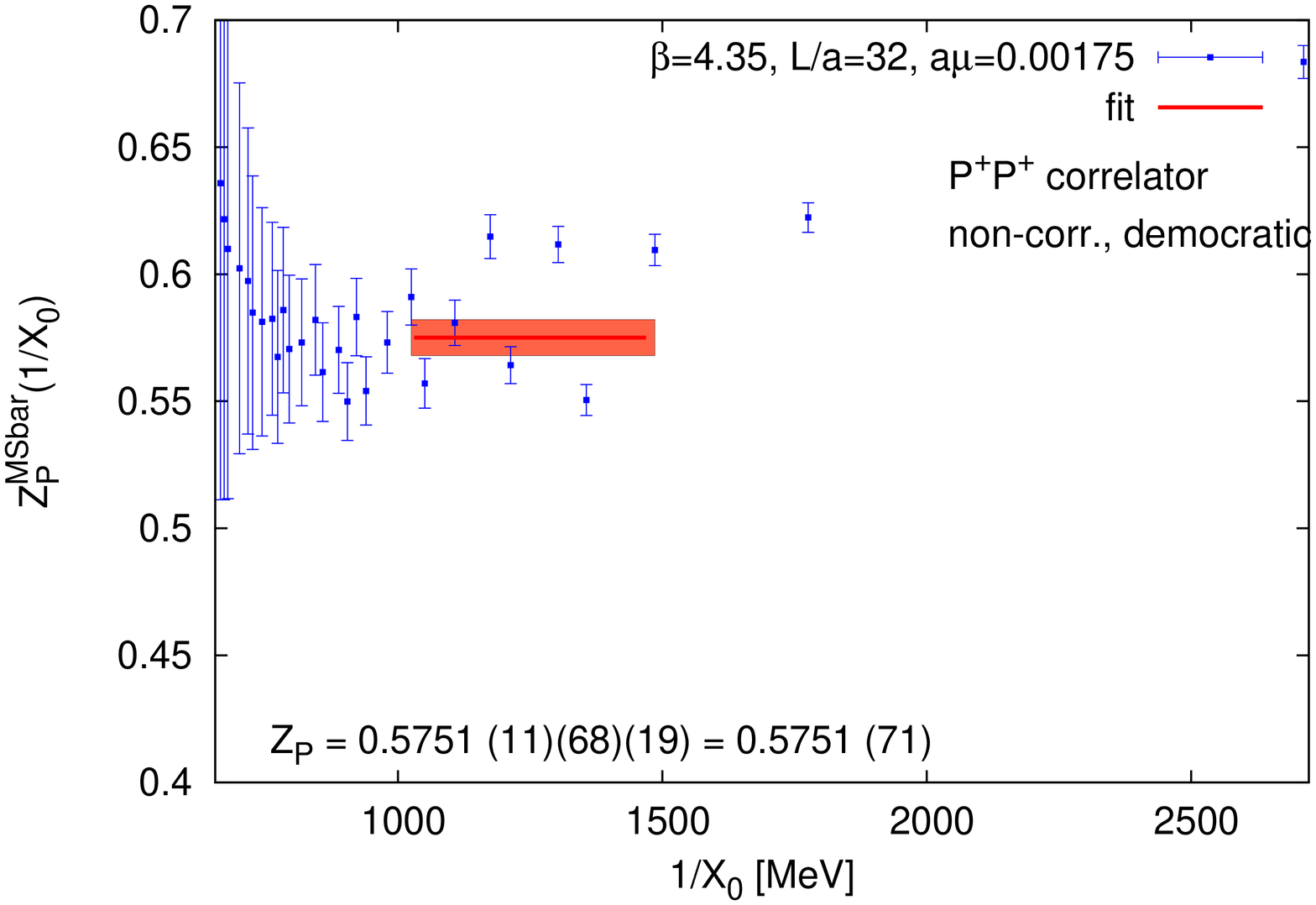}
\includegraphics
[width=0.34\textwidth,angle=270]
{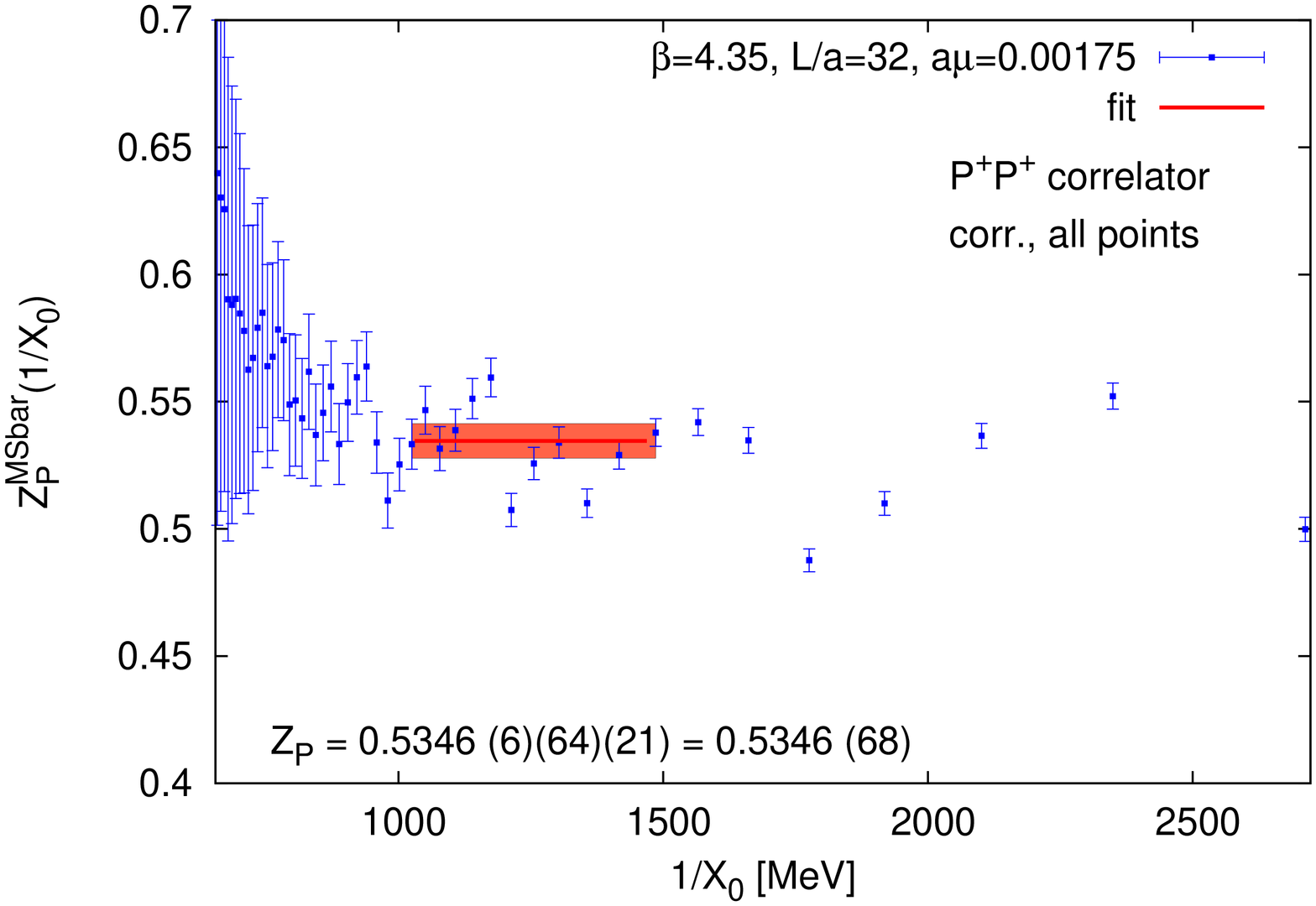}
\includegraphics
[width=0.34\textwidth,angle=270]
{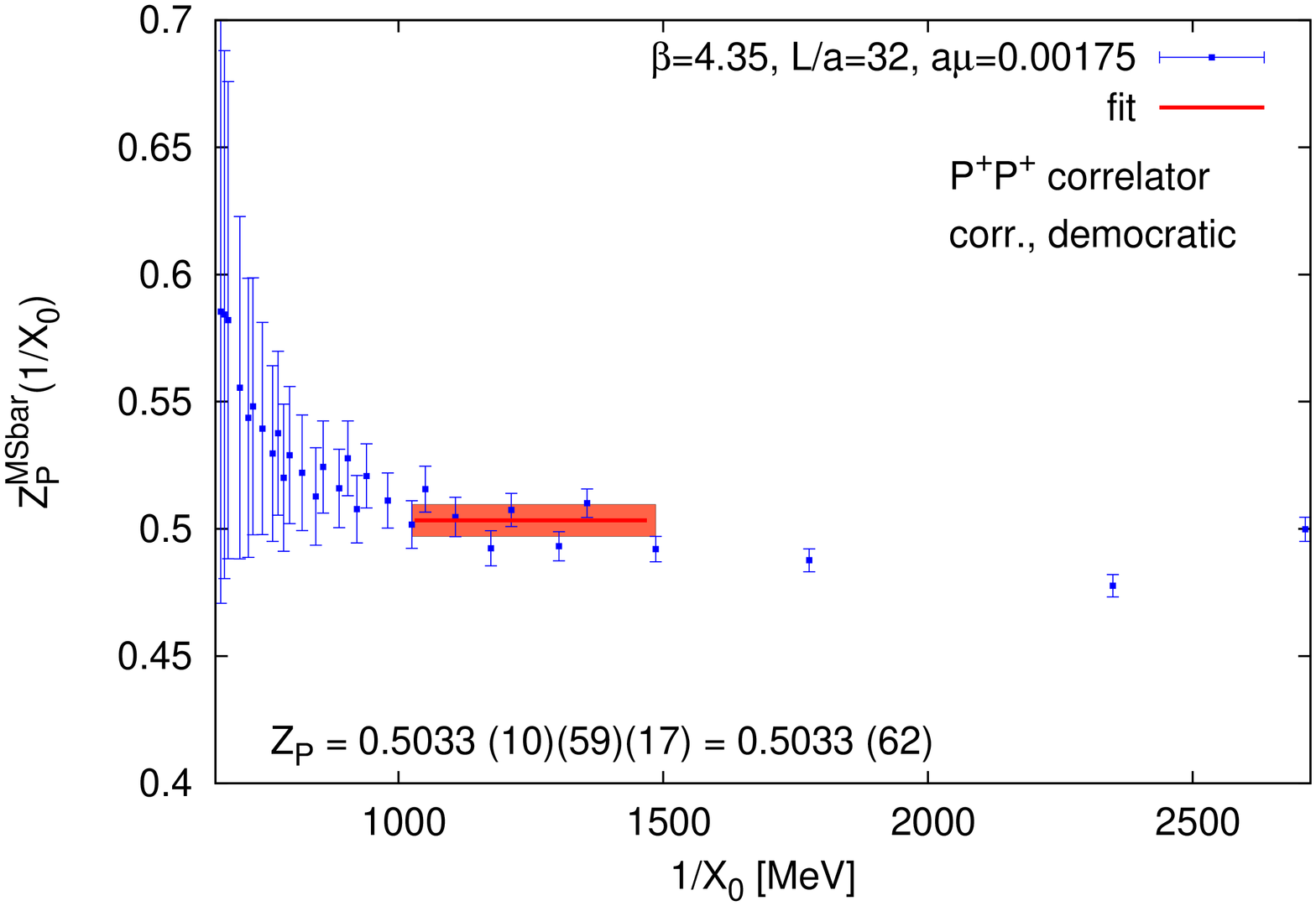}
\end{center}
\caption{\label{fig:ZP_MSbar} The values of RC $Z_P$ in the $\MSb$ scheme at the reference scale of
2 GeV. Ensemble: E17.32. We show the influence of applying tree-level corrections and
``democratic'' cuts on the final values of $Z_{P}^{\MSb}(2\,\mathrm{GeV})$ obtained from the
X-space scheme at different scales $X_0^2$. Upper plots: no tree-level correction applied, lower
plots: with tree-level correction. Left plots: no ``democratic'' cuts (all points), right plots:
only ``democratic'' points. The exclusion of ``non-democratic'' points from the averages results in
vertical shifts of some of the points; hence, the data points shown in the right plots are not
merely subsets of the points shown in the left plots. The final plot of our analysis for $Z_P$ for
this ensemble is the lower right one. Errors as explained in the caption of Tab.~\ref{tab:example}.
Note that the vertical and
horizontal scales are the same in all plots.}
\end{figure}

Our final values for $Z_{P}^{\MSb}(2\,\mathrm{GeV})$ are summarized in
Fig.~\ref{fig:ZP_MSbar}(lower right). The plot shows that within our extraction window, the final
value of $Z_{P}^{\MSb}(2\,\mathrm{GeV})$ does not depend on the choice of the renormalization scale
$1/X_0$ for
the X-space scheme. However, fluctuations around the central value are well visible. As stated
above, the main source of these fluctuations are the remaining non-subtracted cut-off
effects. We
will discuss their role below.

Fig.~\ref{fig:ZP_MSbar}(lower right) also shows the extraction of the final value of
$Z_{P}^{\MSb}(2\,\mathrm{GeV})$, resulting from averaging over different renormalization points in
the X-space scheme. To take the systematic errors appropriately into account, we perform
this averaging in the following way. We take the central values of
$Z_{P}^{\MSb}(2\,\mathrm{GeV})(1/X_0)$ (where this notation shows that they were extracted from
different renormalization points in
the X-space scheme) with their statistical errors and compute the weighted average. The error of
this weighted average is then the statistical error of the final value of
$Z_{P}^{\MSb}(2\,\mathrm{GeV})$.
Next, we compute the weighted averages for the values of $Z_{P}^{\MSb}(2\,\mathrm{GeV})(1/X_0)$
shifted upwards and downwards by the lattice spacing systematic error. The spread of the resulting
weighted averages defines the lattice spacing systematic error of the final value
$Z_{P}^{\MSb}(2\,\mathrm{GeV})$. Finally, we similarly extract the $\Lambda_{\MSb,N_f=2}$ related
systematic error. 
The ultimate error of $Z_{P}^{\MSb}(2\,\mathrm{GeV})$ comes from the statistical and two systematic
errors combined in quadrature.
Such procedure finally gives
$Z_{P}^{\MSb}(\mu=2\,\mathrm{GeV})\,=\,0.5033(10)(59)(17)\,=\,0.503(6)$ for ensemble E17.32.

The different plots of Fig.~\ref{fig:ZP_MSbar} illustrate the role of tree-level corrections and
``democratic'' cuts. The upper left plot corresponds to non-corrected correlators and using all
points. Depending on the choice of $X_0^2$, the X-space RC value converted to the $\MSb$ scheme and
evolved to 2 GeV can yield values that are up to ca. 25\% away. For coarser values of the lattice
spacing, this difference can even exceed 50\%.
Therefore, as we have argued above, the tree-level correction and ``democratic'' cuts are crucial
for a reliable analysis. As the lower left and upper right plots of Fig.~\ref{fig:ZP_MSbar} show,
the correction or cuts exclusively provide a big improvement over the original correlator case.
However, it is obvious that only the two corrections combined can give a really good quality of
the extracted value of $Z_P$.

This concludes our detailed discussion of the extraction of $Z_P$ for ensemble E17.32. In the next
section, we summarize our results for all RCs and all ensembles.

\section{Results and comparison}
\label{sec. analysis2}

\subsection{RCs for ETMC ensembles}
\label{sec. results}

\begin{figure}[t!]
\begin{center}
\includegraphics
[width=0.34\textwidth,angle=270]
{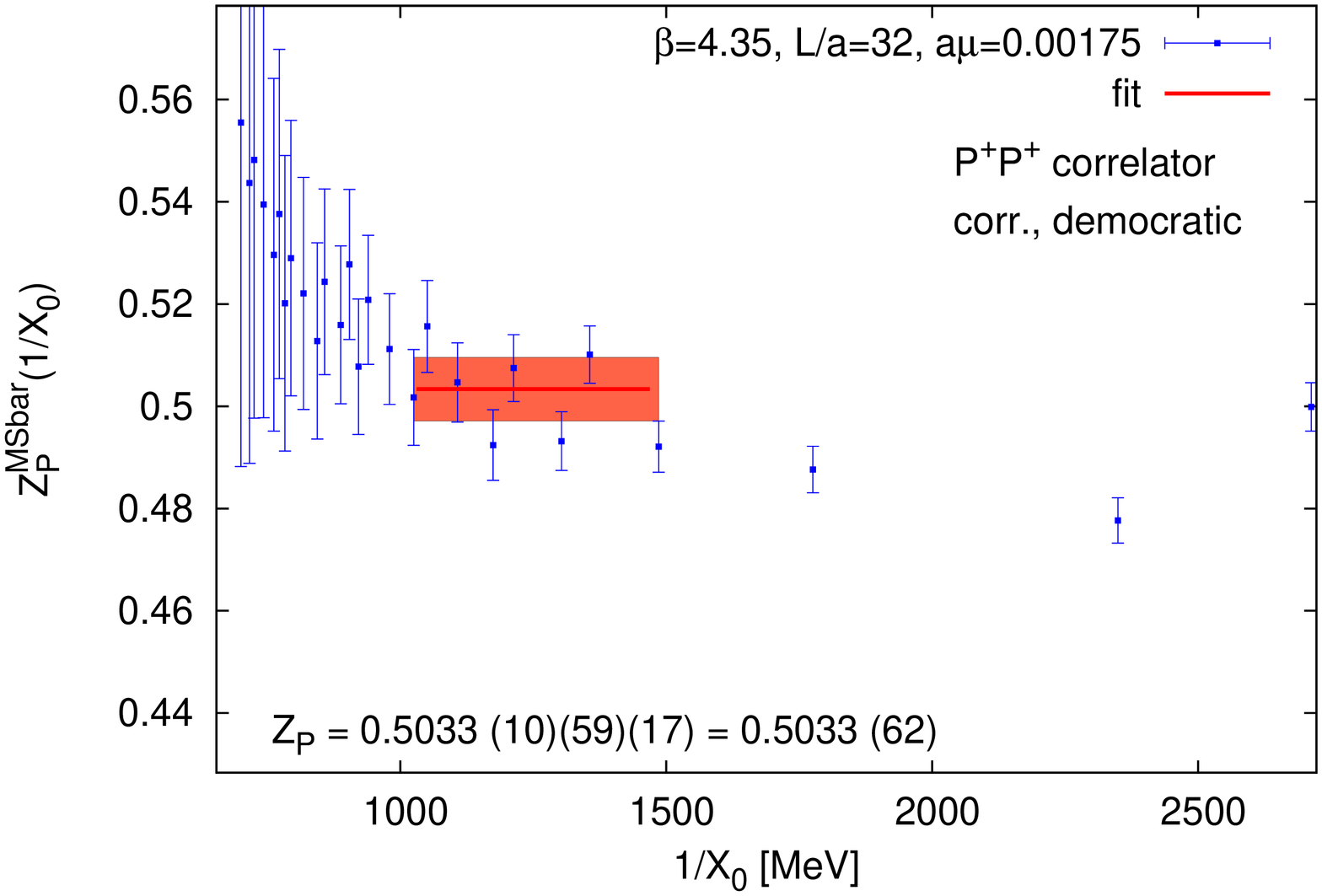}
\includegraphics
[width=0.34\textwidth,angle=270]
{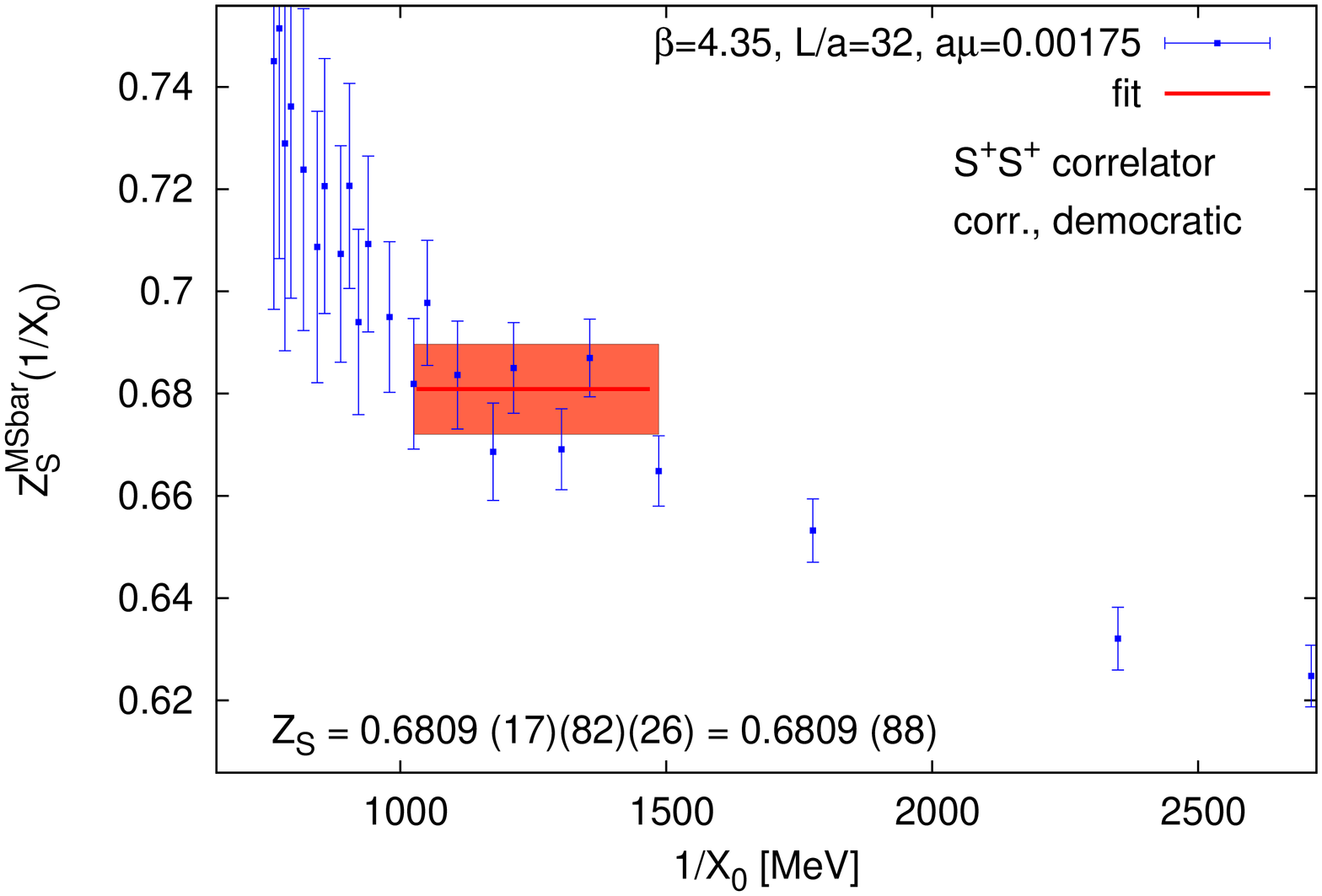}
\includegraphics
[width=0.34\textwidth,angle=270]
{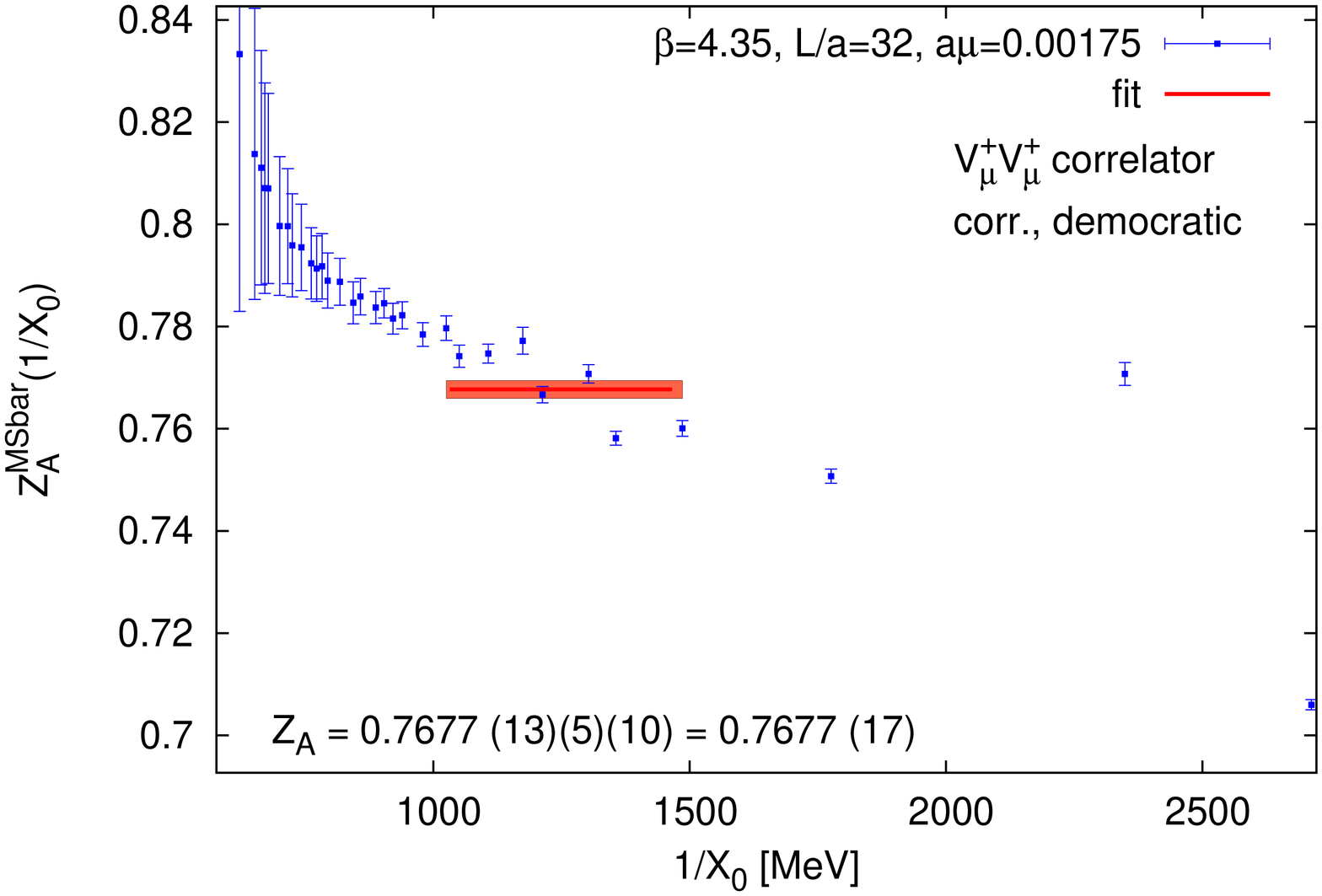}
\includegraphics
[width=0.34\textwidth,angle=270]
{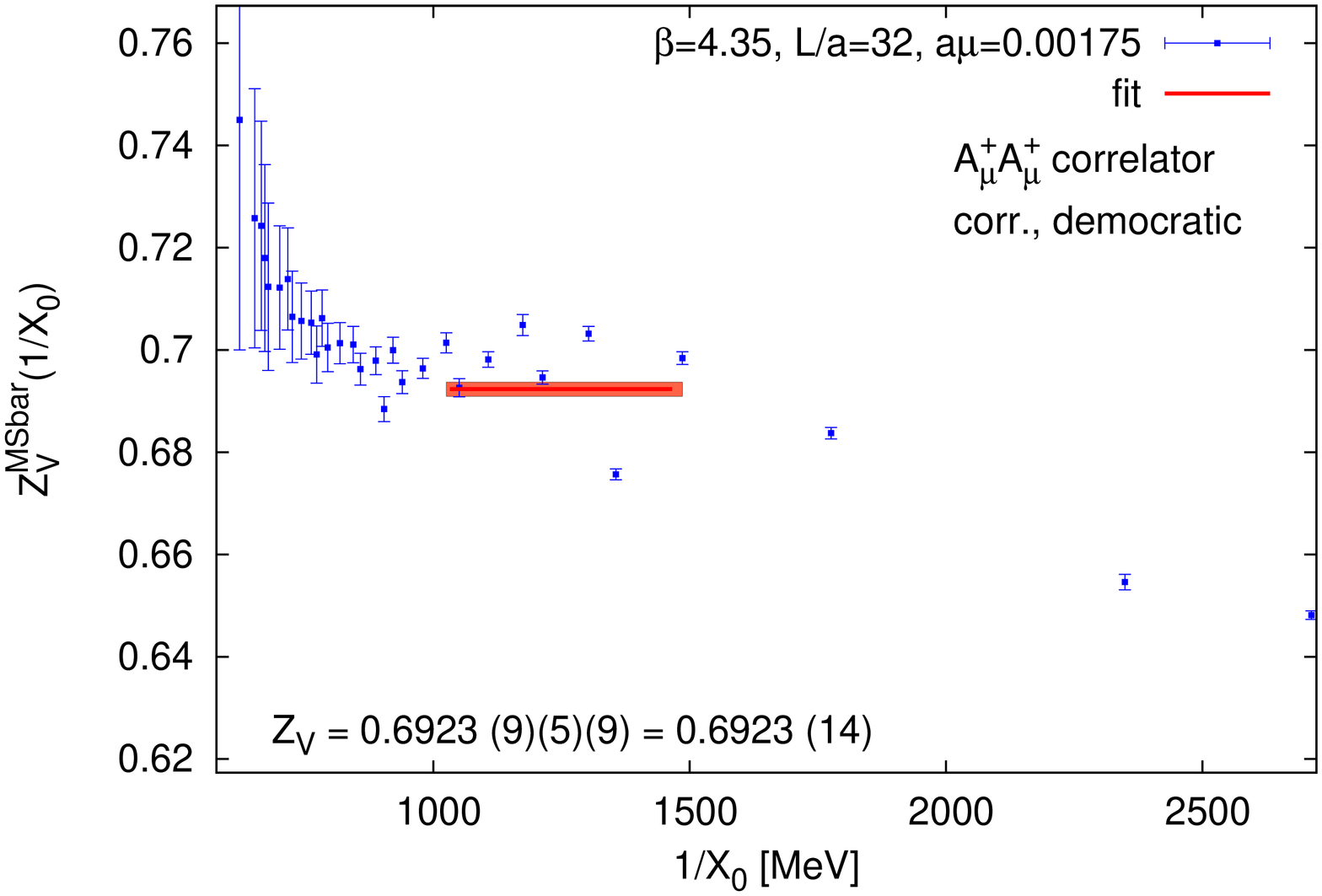}
\includegraphics
[width=0.34\textwidth,angle=270]
{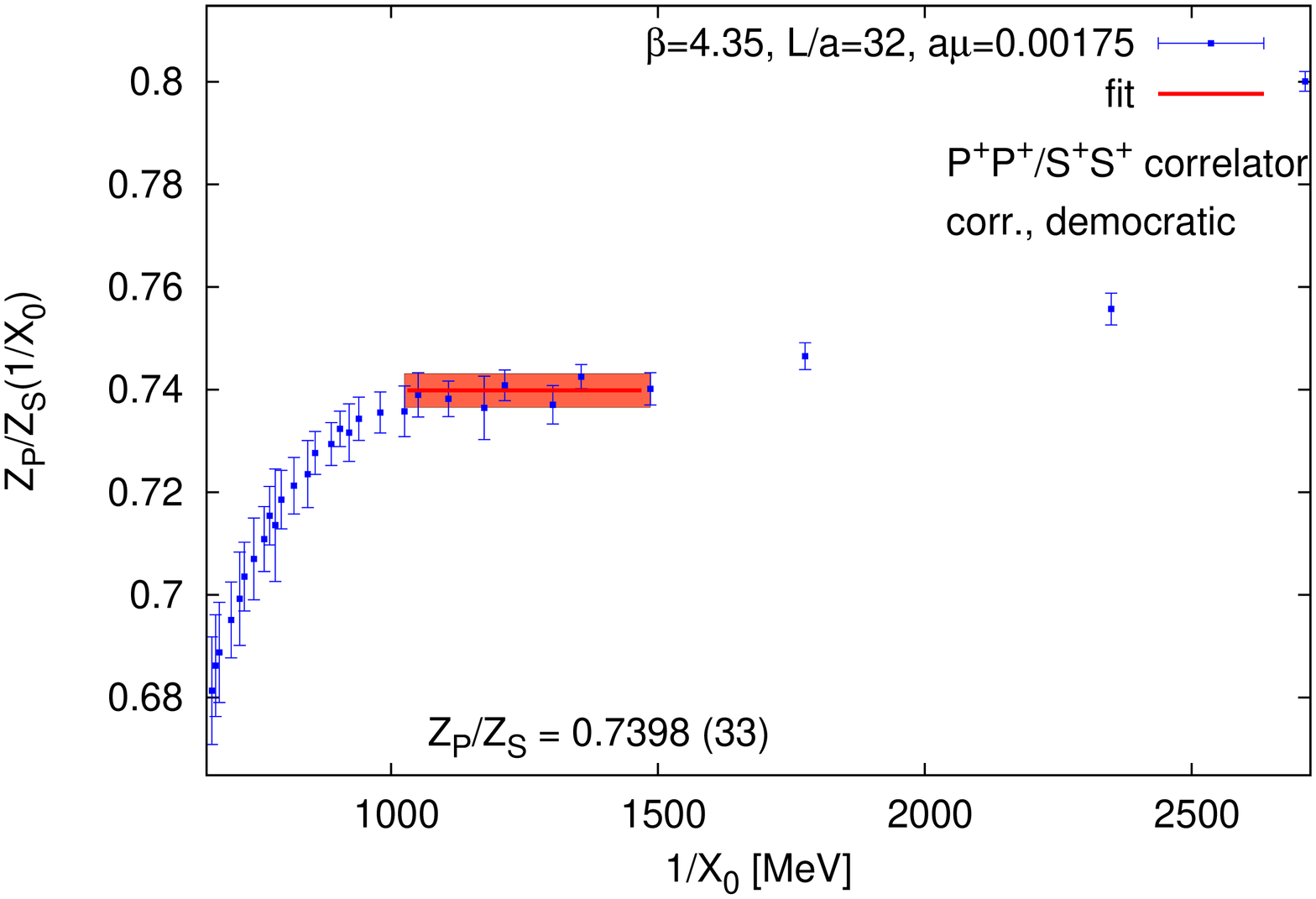}
\includegraphics
[width=0.34\textwidth,angle=270]
{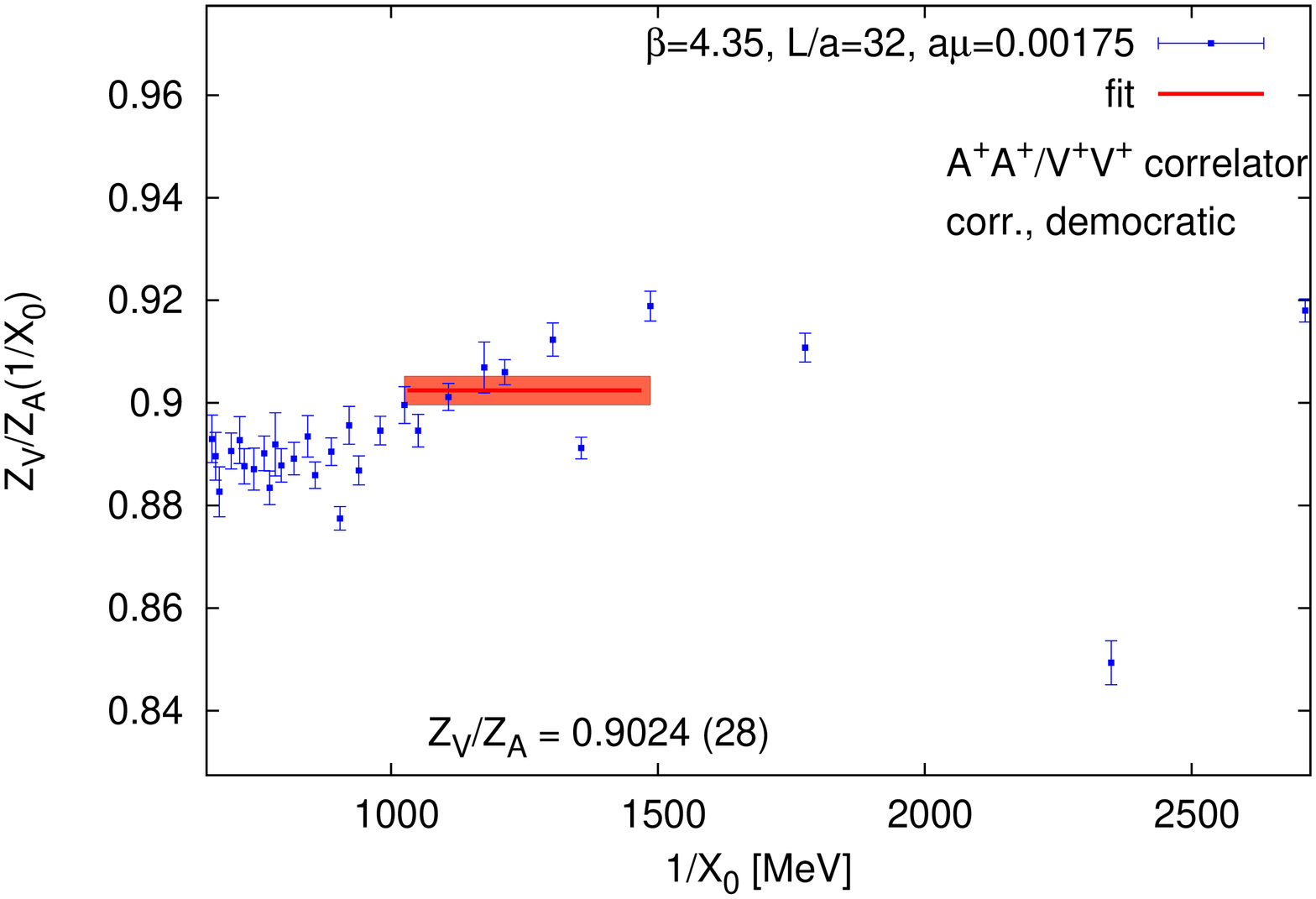}
\end{center}
\caption{\label{fig:all_RCs} Extraction of final values of RCs $Z_P$, $Z_S$, $Z_V$, $Z_A$ and
the ratios $Z_P/Z_S$, $Z_V/Z_A$ for ensemble E17.32. Errors as explained in the caption of
Tab.~\ref{tab:example}. Note that the horizontal scale is the same in all plots. The vertical
scale is always $[Z_{\Gamma}-0.075,\,Z_{\Gamma}+0.075]$, where $Z_{\Gamma}$ is the extracted value
of RC.} 
\end{figure}

The procedure for other renormalization constants follows the line of the detailed example of the
previous section. To compare the quality of the data for different types of RCs, we plot in
Fig.~\ref{fig:all_RCs} the dependence of $Z_{\Gamma}^{\MSb}(2\,\mathrm{GeV})$ on the X-space
scheme renormalization point. We show the results for $Z_P$, $Z_S$, $Z_V$, $Z_A$ and
the ratios $Z_P/Z_S$, $Z_V/Z_A$ for ensemble E17.32. The errors for $Z_P$ and $Z_S$ are much larger
than the ones for $Z_V$ and $Z_A$, since the latter are scale-independent in the $\MSb$ scheme and
hence do not require the step of evolution to the scale 2 GeV. For the ratios $Z_P/Z_S$ and
$Z_V/Z_A$, we quote only the statistical error, because both the RC in the numerator and the
denominator have the same conversion factor and the same scale dependence -- hence, the conversion
factors cancel each other and similarly for the scale dependence. Thus, $Z_P/Z_S$ and $Z_V/Z_A$ are
fully scale- and scheme-independent\footnote{The RCs $Z_V$ and $Z_A$ are often said to be scale- and
scheme-independent. However, this concerns only renormalization schemes that preserve chiral Ward
identities.}.

\begin{figure}[t!]
\begin{center}
\includegraphics
[width=0.34\textwidth,angle=270]
{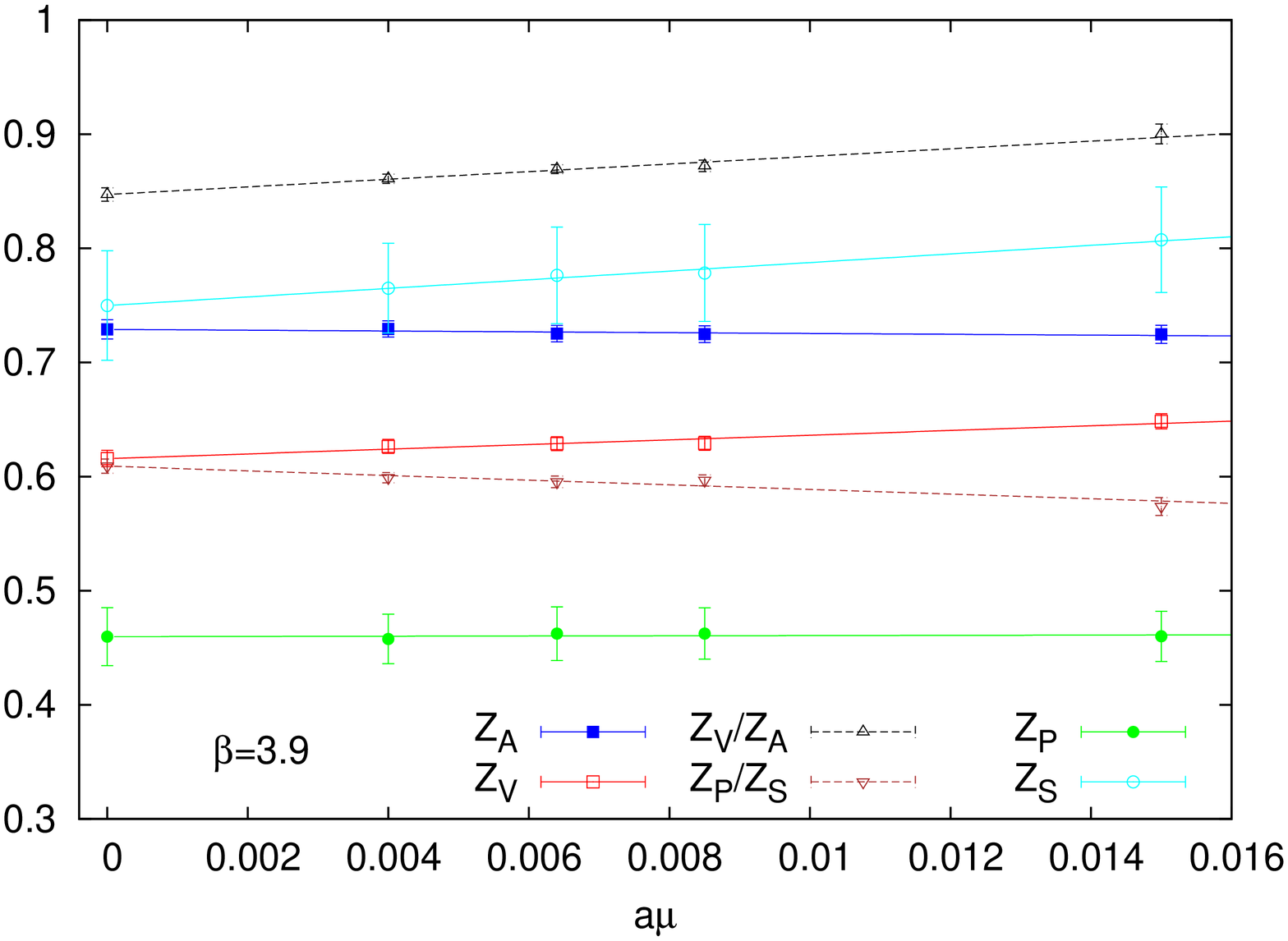}
\includegraphics
[width=0.34\textwidth,angle=270]
{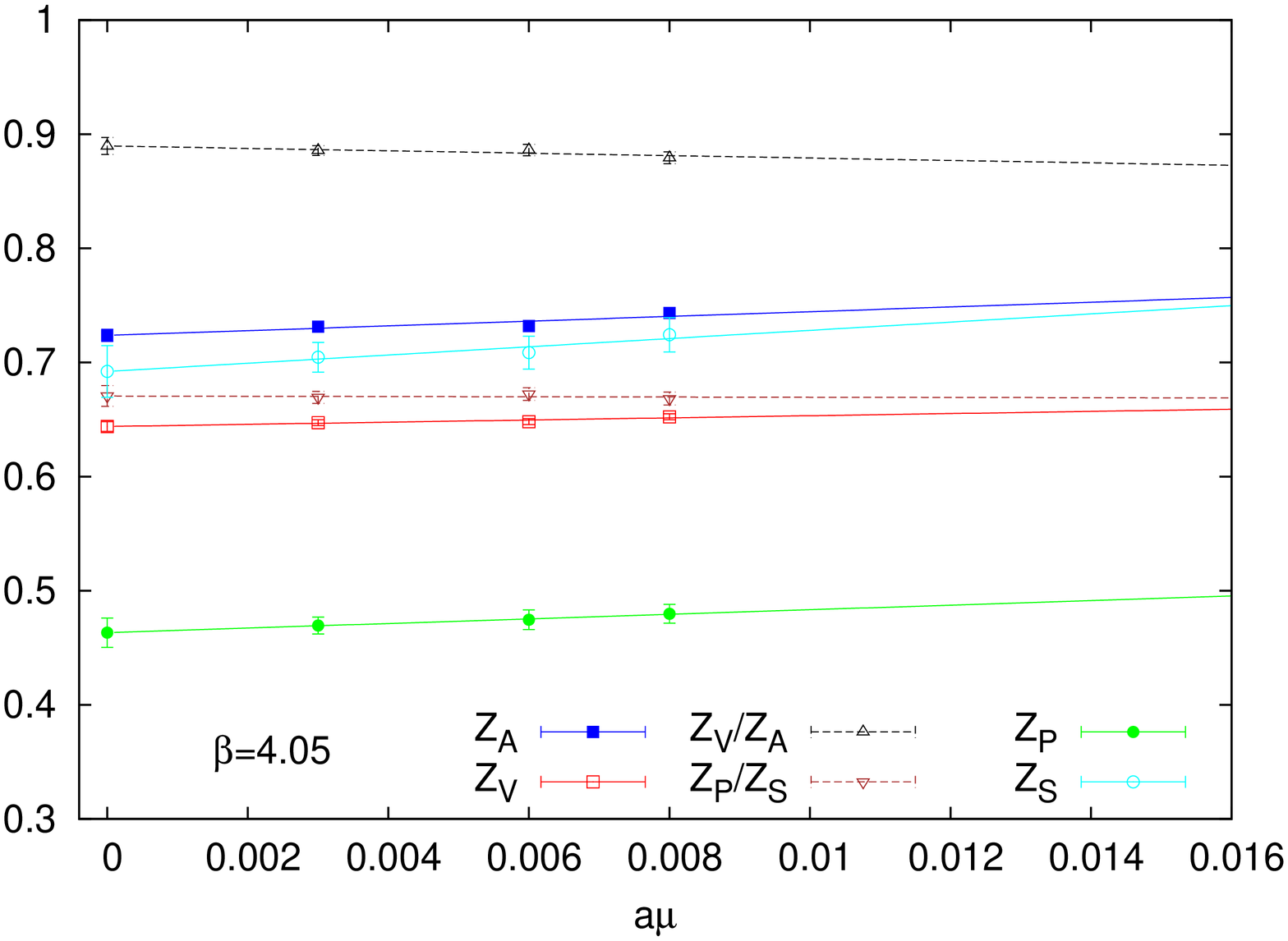}
\caption{\label{fig:chiral} Chiral extrapolations of renormalization constants for
$\beta=3.9$ (left) and $\beta=4.05$ (right).} 
\end{center}
\end{figure}

We now move on to discuss our results for other ETMC ensembles of gauge field configurations. In
particular, for two values of the lattice spacing, we can perform an extrapolation of RCs to the
chiral limit and obtain their mass-independent values. Such extrapolations for $\beta=3.9$
and $4.05$ ensembles are shown in Fig.~\ref{fig:chiral}. In all cases, our results are compatible
with a linear dependence on the quark mass and this dependence is rather mild (in some cases the
slope is compatible with zero). We also observe that the chiral limit value is always compatible
with the value at the lowest quark mass. Therefore, at $\beta=4.2$ and $\beta=4.35$, where only one
ensemble is available to extract X-space RCs, it is plausible to use our extracted values as
mass-independent values, since they correspond to the same pion mass as the lowest quark masses at
$\beta=3.9$ and $4.05$.

\begin{table}[t!]
\begin{center}
%\hspace*{-0.7cm}
\begin{scriptsize}
\begin{tabular}{@{\extracolsep{-2.5mm}} ccccccc}
%begin{tabular}[b]{\extracolsep{-5mm}}{ccccccc}
\hline
\hline 
 & $Z_P$ & $Z_S$ & $Z_V$ & $Z_A$ & $Z_P/Z_S$ & $Z_V/Z_A$\\
\hline
B40.24 & 0.458(2)(12)(21) &  0.765(3)(21)(38) & 0.626(1)(2)(6) &
0.729(2)(3)(7)& 0.599(4) & 0.861(4)\\ 
B64.24 & 0.462(2)(13)(23) & 0.776(3)(22)(41) & 0.629(1)(2)(6) &
0.725(2)(3)(7) & 0.595(5) & 0.869(4)\\
B85.24 & 0.463(2)(12)(22) & 0.778(4)(22)(41) & 0.629(2)(2)(6) &
0.725(2)(3)(7) & 0.597(5) & 0.872(5)\\
B150.24 & 0.460(2)(12)(21) & 0.807(7)(24)(45) & 0.648(3)(2)(6) &
0.725(4)(3)(7) & 0.574(8) & 0.900(9)\\
chiral & 0.460(2)(14)(25) &  0.750(5)(25)(47) &
 0.616(2)(3)(7) &
 0.729(3)(3)(8) &  0.609(6) &  0.847(6)\\
\cite{Constantinou:2010gr} & 0.437(7) &  0.713(10) &  0.624(4) &  0.746(6) &
 0.613(13) &  0.836(12)\\
\cite{Alexandrou:2012mt} &  0.457(10)(16) &  0.726(5)(11) &  0.627(1)(3) &
  0.758(1)(1) &  0.639(3)(1) &  0.827(2)(5) \\ 
\hline
C30.32 & 0.469(1)(6)(7) & 0.705(3)(10)(12) & 0.647(1)(1)(2) &
0.731(2)(1)(2) & 0.669(5) & 0.886(4)\\
C60.32 & 0.475(2)(7)(8) & 0.709(3)(11)(13) & 0.648(2)(1)(2) &
0.732(2)(1)(2) & 0.672(6) & 0.886(5)\\
C80.32 & 0.480(2)(7)(7) & 0.724(4)(11)(14) & 0.652(2)(1)(2) &
0.743(3)(1)(2) & 0.668(6) & 0.879(5)\\
chiral &  0.463(2)(11)(11) &  0.692(6)(17)(20) &  0.644(2)(1)(3) &
 0.724(3)(1)(4) &  0.671(9) &  0.890(8)\\
\cite{Constantinou:2010gr} &  0.477(6) &  0.699(6) &  0.659(3) &  0.772(6) &
 0.682(12) &  0.854(10)\\
\cite{Alexandrou:2012mt} &  0.497(8)(15) &  0.691(9)(16) &  0.662(1)(3) &
 0.773(1)(1) &  0.682(2)(1) &  0.856(2)(5)\\
\hline
D20.48 &  0.516(5)(3)(5) &  0.730(7)(4)(7) &  0.702(6)(1)(1) &
 0.785(7)(1)(2) &  0.707(14) &  0.893(16)\\
\cite{Alexandrou:2012mt} &  0.501(8)(10) &  0.695(10)(13) &  0.686(1)(1) &
 0.789(1)(2) &  0.713(2)(2) &  0.869(2)(4)\\
\hline
E17.32 & 0.503(1)(6)(2) & 0.681(2)(8)(3) & 0.692(1)(1)(1) &
0.768(1)(1)(1) & 0.740(3) & 0.902(3)\\
\hline
\hline
\end{tabular}
\end{scriptsize}
\end{center}
\caption{\label{tab:results} Renormalization constants for all ensembles, extracted in the X-space
renormalization scheme, converted to the $\MSb$ scheme and evolved to the reference scale of 2 GeV
for scale-dependent RCs. For $\beta=3.9$ and $4.05$, we show our values in the chiral limit. For
comparison, we also give values determined in the RI-MOM scheme \cite{Constantinou:2010gr},
\cite{Alexandrou:2012mt}. Errors are:  1st -- statistical, 2nd -- resulting
from the uncertainty in lattice
spacing, 3rd -- resulting from the uncertainty of $\Lambda_{\MSb,N_f=2}$.
The last two errors do not apply for $Z_P/Z_S$ and $Z_V/Z_A$, see text.}
\end{table} 

Tab.~\ref{tab:results} summarizes all our final results for the renormalization constants extracted
in the X-space renormalization scheme, together with the values determined in the RI-MOM scheme,
given for comparison \cite{Constantinou:2010gr}, \cite{Alexandrou:2012mt}.
Although the cut-off effects in the X-space method and in RI-MOM are, in principle, very different,
in most cases we find compatible values. This strengthens the expectation 
that both methods should lead to only rather mild
discretization effects. The errors of the X-space method are in general slightly larger than the
ones of RI-MOM. This is particularly visible at $\beta=3.9$, where the by far dominating error is
related to the necessity of going rather close to $\Lambda_{\MSb,N_f=2}$ (down to around 800 MeV),
where the strong coupling constant and hence the conversion factors X-space$\,\rightarrow\MSb$
strongly depend on the actual value of $\Lambda_{\MSb,N_f=2}$. This uncertainty becomes less
important at finer lattice spacings and then in some cases we can achieve comparable or even
smaller errors than RI-MOM (e.g. $Z_P$, $Z_S$ at $\beta=4.2$). In general, our precision in $Z_P$,
$Z_S$ increases from 5-7\% at the coarsest lattice spacing to better than 1\% at its smallest
value. For $Z_V$, $Z_A$ it is approximately 1\% already at $\beta=3.9$ and it reaches even 1-2 per
mille at
$\beta=4.35$. We observe a similar behaviour for the ratios $Z_P/Z_S$ and $Z_V/Z_A$\footnote{Note
that $\beta=4.2$ is an exception to these observations. This is only due to statistical accuracy --
for the ensemble D20.48 we have only limited statistics.}.

\subsection{$\Oag$ effects}
\label{sec. Oa2g2}

\begin{table}[t!]
\begin{center}
%\hspace*{-0.7cm}
\begin{scriptsize}
\begin{tabular}{@{\extracolsep{-2.5mm}} ccccccc}
%begin{tabular}[b]{\extracolsep{-5mm}}{ccccccc}
\hline
\hline 
 & $Z_P$ & $Z_S$ & $Z_V$ & $Z_A$ & $Z_P/Z_S$ & $Z_V/Z_A$\\
\hline
B40.24 & 0.458(24)(25) &  0.765(44)(73) & 0.626(6)(31) &
0.729(8)(15)& 0.599(4)(35) & 0.861(4)(25)\\ 
B64.24 & 0.462(26)(26) & 0.776(47)(78) & 0.629(6)(33) &
0.725(8)(14) & 0.595(5)(36) & 0.869(4)(28)\\
B85.24 & 0.463(25)(27) & 0.778(47)(76) & 0.629(6)(34) &
0.725(8)(14) & 0.597(5)(33) & 0.872(5)(33)\\
B150.24 & 0.460(25)(26) & 0.807(51)(90) & 0.648(7)(43) &
0.725(8)(14) & 0.574(8)(41) & 0.900(9)(41)\\
chiral &  0.460(29)(29) &  0.750(53)(90) &
 0.616(8)(40) &
 0.729(9)(17) &  0.609(6)(42) &  0.847(6)(34)\\
\hline
C30.32 & 0.469(9)(13) & 0.705(16)(40) & 0.647(3)(19) &
0.731(3)(12) & 0.669(5)(20) & 0.886(4)(17)\\
C60.32 & 0.475(10)(15) & 0.709(17)(39) & 0.648(3)(20) &
0.732(3)(11) & 0.672(6)(15) & 0.886(5)(15)\\
C80.32 & 0.480(10)(18) & 0.724(18)(42) & 0.652(3)(23) &
0.743(4)(11) & 0.668(6)(14) & 0.879(5)(19)\\
chiral &  0.463(16)(23) &  0.692(27)(68) &  0.644(4)(34) &
0.724(5)(19) &  0.671(9)(31) &  0.890(8)(29)\\
\hline
D20.48 &  0.516(8)(10) &  0.730(11)(17) &  0.702(6)(17) &
 0.785(7)(10) &  0.707(14)(8) &  0.893(16)(13)\\
\hline
E17.32 & 0.503(6)(12) & 0.681(9)(16) & 0.692(1)(15) &
0.768(2)(11) & 0.740(3)(3) & 0.902(3)(14)\\
\hline
\hline
\end{tabular}
\end{scriptsize}
\end{center}
\caption{\label{tab:results2} Comparison of the total error for each RC extracted in X-space (the
number in 1st parentheses) and the magnitude of fluctuations of RCs in the extraction window (2nd
parentheses), due to the un-subtracted cut-off effects ($\Oag$ at leading order). The
total error is the statistical error and two systematic errors, combined in quadrature.}
\end{table} 

As we have discussed above, our analysis procedure involves the subtraction of tree-level
discretization effects and ``democratic'' cuts, which eliminate points that are subject to enhanced
discretization effects due to the breaking of rotational symmetry.
However, we have not eliminated any of the leading order of lattice perturbation 
theory $\Oag$ effects. As was
already suggested in Ref.~\cite{Gimenez:2004me}, the fluctuations of RCs observed in the extraction
window $a\ll X_0 \ll\Lambda^{-1}_{\mathrm{QCD}}$ can be due mainly to these cut-off effects (and
higher
order ones).
In other words, if true, subtracting these effects as a part of the analysis procedure should lead to
compatible values of RCs for all choices of the renormalization point $X_0$.
In our study, we found further evidence that the fluctuations are indeed caused by un-subtracted
discretization effects -- the foremost being that their magnitude clearly diminishes with
decreasing lattice spacing. Moreover, the effects seem to be rather regular, when one compares the
same $Z_\Gamma$ for different values of $\beta$ -- suggesting that these are not random
fluctuations, but some systematic effects.

Our estimates of the size of these fluctuations is shown in Tab.~\ref{tab:results2}, which shows
the extracted values of RCs, together with their total errors (statistical + two systematic errors
combined in quadrature) -- the number in the first parentheses, and one half of the spread between
the highest estimate of $Z_\Gamma^{\MSb}(2\,\mathrm{GeV})$ in the extraction window and its lowest
estimate -- the number in the second parentheses.
In general, the magnitude of fluctuations is 1-15 times the total error, with a systematic
tendency to decrease for finer lattice spacings.

A systematic way to improve the X-space method would be to compute these effects in lattice
perturbation theory or numerical stochastic perturbation theory. This has the potential 
to reduce the size of
fluctuations and hence enhance the range of applicability and the reliability of the method. Our
present approach averages the fluctuations out. However, since, $\Oag$ effects
can be small for some types of points and much larger for some others, it will be 
important to understand the size of these perturbative corrections in detail.

\section{Conclusions}
\label{sec. conclusions}

In this paper, we presented a feasibility study 
for the application of the fully gauge invariant X-space renormalization scheme
method to extract renormalization constants of bilinear, flavour
non-singlet operators. To this end, we have used ETMC $N_f=2$ ensembles of maximally twisted mass
fermions on a tree-level Symanzik improved gauge action. Using continuum perturbation theory
formulae to convert between the X-space scheme and the $\MSb$ scheme, we also computed the values
of RCs in the latter, at a reference scale of 2 GeV.
This allowed us to compare our results to the ones obtained in the RI-MOM scheme and likewise
converted to $\MSb$.
In general, we found very good agreement between the two non-perturbative schemes, signaling rather
small cut-off effects in both approaches.
For our coarsest lattice spacings, our errors are visibly larger than the ones of the RI-MOM
scheme. However, going to finer lattice spacings, the errors become comparable in size.
Hence, as expected, we conclude that the X-space method works best for fine lattice spacings, where
the renormalization window, the region of small distances satisfying the constraint $a\ll X_0
\ll\Lambda^{-1}_{\mathrm{QCD}}$, becomes sufficiently wide and allows for a reliable extraction of
RCs.
In practical terms, the lattice spacing of approximately 0.08 fm (our $\beta=3.9$ ensembles) is on
the
verge of applicability of the method.

The X-space method was successfully used before for the computation of RCs in the quenched case
with the standard Wilson gluonic action \cite{Gimenez:2004me}, while in this paper, we work in a
dynamical setup. Apart from this, the main differences with respect to Ref.~\cite{Gimenez:2004me}
are: the use of ``democratic'' points to reduce cut-off effects and 4-loop conversion formulae
between the X-space and $\MSb$ schemes. Both seem to be important ingredients for the
reliability of the X-space method.
Further improvement can presumably be obtained by calculating $\Oag$ effects in LPT or NSPT and
subtracting
them from the correlators. Thus, the values of RCs would be even less contaminated by cut-off
effects.

To finalize, let us summarize the main features of the X-space method. Being fully gauge
invariant and free of contact terms, it prevents mixing with certain types of operators. This makes
it suitable for the computation of RCs needed to renormalize weak matrix elements. This, however, requires the
additional step of deriving continuum perturbation theory expressions relating the X-space RCs to
the $\MSb$ ones. The X-space method has also some practically appealing features -- it is easy to
implement and relatively cheap computationally.
Thus, we believe that the X-space renormalization scheme has promising potential as a 
method to perform non-perturbative renormalization in lattice QCD in the future.

\vspace{0.3cm}
\noindent {\bf Acknowledgments} We thank M.~Petschlies for computing correlation
functions in coordinate space and making them available to us. We also thank the whole European
Twisted Mass Collaboration for the collective effort of generating ensembles of gauge field
configurations that we have used for this work. We are especially grateful to V.~Lubicz for several
illuminating discussions about the X-space method and to F.~Sanfilippo for carefully
reading the manuscript and several useful comments and suggestions. We also acknowledge valuable
discussions and
suggestions from K.~G.~Chetyrkin, R.~Frezzotti, G.~Herdoiza, J.~H.~K\"uhn, M.~Petschlies
and R.~Sommer. 
K.C. has been supported by Foundation for Polish Science fellowship ``Kolumb''.
This work has been supported in part by the DFG Sonderforschungsbereich/Transregio SFB/TR9.

\end{document}